\documentclass[11pt, twocolumn]{article}
\usepackage[normalem]{ulem}
\usepackage{authblk}
\usepackage{graphicx}
\usepackage{float}
\usepackage{xcolor}
\usepackage{amsmath}
\usepackage{natbib}
\usepackage{units}
\usepackage[margin=0.5in]{geometry}
\RequirePackage{fixltx2e}
\usepackage[font=small,labelfont=bf]{caption}
\usepackage{float}
\begin{document}
\title{\vspace{0cm}Tests for coronal electron temperature signatures in suprathermal electron populations at 1 AU}
\author[1]{Allan R. Macneil}
\author[1]{Christopher J. Owen}
\author[1,2]{Robert T. Wicks}
\affil[1]{Mullard Space Science Laboratory, University College London, Surrey, U.K.}
\affil[2]{Institute for Risk and Disaster Reduction, University College London, London, U.K.}

%\affil[1]{Mullard Space Science Laboratory, University College London, Surrey, U.K.}
%\affil[2]{Institute for Risk and Disaster Reduction, University College London, London, U.K.}
%
%\runningtitle{Tests for coronal electron temperature signatures}

%\correspondence{A.\,R.\,Macneil: allan.macneil.15@ucl.ac.uk}

%\received{}
%\pubdiscuss{} 
%\revised{}
%\accepted{}
%\published{}

%% These dates will be inserted by Copernicus Publications during the typesetting process.

%\firstpage{1}

\twocolumn[
  \begin{@twocolumnfalse}
    \maketitle
\begin{abstract}
The development of knowledge of how the coronal origin of the solar wind affects its in situ properties is one of the keys to understanding the relationship between the Sun and the heliosphere. 
In this paper, we analyse ACE/SWICS and WIND/3DP data spanning $>$12 years, and test properties of solar wind suprathermal electron distributions for the presence of signatures of the coronal temperature at their origin which may remain at 1\,AU. In particular we re-examine a previous suggestion that these properties correlate with the oxygen charge state ratio $\text{O}^{7{+}}{\negmedspace/}\text{O}^{6{+}}$; an established proxy for coronal electron temperature. We find only a very weak but variable correlation between measures of suprathermal electron energy content and $\text{O}^{7{+}}{\negmedspace/}\text{O}^{6{+}}$. The weak nature of the correlation leads us to conclude, in contrast to earlier results,
that an initial relationship with core electron temperature has the possibility to exist in the corona, but that in most cases no strong signatures remain in the suprathermal electron distributions at 1\,AU. 
It can not yet be confirmed whether this is due to the effects of coronal conditions on the establishment of this relationship, or to the altering of the electron distributions by processing during transport in the solar wind en route to 1\,AU.
Contrasting results for the halo and strahl population favours the latter interpretation. Confirmation of this will be possible using Solar Orbiter data (cruise and nominal mission phase) to test whether the weakness of the relationship persists over a range of heliocentric distances. If the correlation is found to strengthen when closer  to the Sun, then this would indicate an initial relationship which is being degraded, perhaps by wave-particle interactions, en route to the observer. 
\vspace{11pt}
\end{abstract}
  \end{@twocolumnfalse}
]

\section{Introduction}
\label{sec:intro} 
Solar wind plasma populations leaving the Sun can be  expected to have properties that reflect conditions of their source regions. 
However, during the course of the wind's propagation out to 1\,AU and beyond, internal dynamic processes may develop within  the solar wind plasma.
These cause many of the solar wind properties to be altered to the extent that the signatures of their solar source, such as proton temperature \citep{Freeman1988} and bulk speed \citep{Schwenn1990}, are no longer clear. 
Nevertheless, the degree of ionisation of heavy ion species in the solar wind provides a well-established means by which the temperature of its coronal source may be inferred, even when observed at 1\,AU \citep{Hundhausen1968a}. 

Solar wind heavy ion populations are frequently characterised through metrics such as statistical abundance ratios between charge states of a given ion (e.g.\ $\text{O}^{7{+}}{\negmedspace/}\text{O}^{6{+}}$) and the mean charge of all measured ions of a given species \citep[see the review article by][for details]{Bochsler2007}. 
\cite{Hundhausen1968a} first predicted that this ionisation information could allow estimates of the coronal electron temperature. The fraction of ions which exist in a given charge state in the corona is determined by the equilibrium between 
the dominant processes of  collisional ionisation and 
radiative   recombination. 
The resulting distribution of an ion population into its charge states is thus a function of electron temperature. 
For a solar wind sample, a derived ``freeze-in temperature'' is indicative of the thermal temperature of the  electron population at the location in the corona where the density falls below a theoretical critical value for ionisation and recombination to effectively cease. At this point the ions become ``frozen-in'' to their charge states. As density drops-off radially, this critical density is associated with a characteristic ``freezing height'' \citep{Owocki1983}, 
which is different for each ion. \cite{Ko1997} estimated that these heights should lie between 1.5--4\,R$_{\odot}$; low in the corona. 
 As electron density continues to fall off with radial distance out in the solar wind,  ion charge states are preserved along magnetic field lines out into the solar wind, as long as the solar wind plasma fulfills the frozen-in flux condition. This then provides a snapshot of coronal electron temperature which can be measured in situ. Given the ratio between two subsequent states of ionisation, $n_i$ and $n_{i+1}$ can be expressed as the ratio between the rates of  collisional ionisation out of state $n_i$; $C_i$ and the rate of recombination out of state $n_{i+1}$; $R_{i+1}$: 
\begin{equation}\label{eq:ratio}
\frac{n_{i+1}}{n_i} = \frac{C_i(T_e)}{R_{i+1}(T_e)}
\end{equation}
This applies only for the case in which the rate of coronal expansion is slow compared to the rate at which ionisation can equilibriate. 
From Equation \ref{eq:ratio} the freeze-in temperature can be estimated from a given adjacent charge state ratio. 

Typically coronal holes, which are thought to produce fast solar wind, are cooler
in electron temperature  than their closed-field counterparts which are generally 
associated with  the 
 slow solar wind \citep{Zirker1977}. The coronal hole wind (CHW) thus typically features lower ionisation states than non-coronal hole wind (NCHW), and plotting solar wind speed alongside, for example, $\text{O}^{7{+}}{\negmedspace/}\text{O}^{6{+}}$, will show a clear anti-correlation \cite[see for example Fig.\,1 in][]{Zurbuchen2012}. As the charge states are frozen-in at a few solar radii, boundaries between streams in the solar wind from different source regions should be preserved in composition data. Indeed as energy and momentum may be transferred across such boundaries, 
their structure should be better preserved in composition than, say, flow velocity. However,  boundaries in ionisation state measurements are sometimes observed to be smoothed out at the trailing edge of solar wind streams, in a similar manner to velocity \citep{Schwadron2005a,Borovsky2016}.
 
\cite{Ko2014} found $\text{O}^{7{+}}{\negmedspace/}\text{O}^{6{+}}$ to be a better tracer of coronal origin than velocity observationally. They report that wind from both the northern and southern polar coronal holes could exhibit the same wind speed, but have a notable discrepancy in their corresponding $\text{O}^{7{+}}{\negmedspace/}\text{O}^{6{+}}$ values. This indicates that $\text{O}^{7{+}}{\negmedspace/}\text{O}^{6{+}}$  is a feature characteristic of the source region, providing information which is not available from velocity data alone. 

Solar wind electrons are commonly described as consisting of three distinct populations \citep{Pierrard2001}: a thermal core, anisotropised by the magnetic field; a near-isotropic  and suprathermal halo; and a strongly field-aligned suprathermal strahl. 
A more energetic and even less dense fourth population, dubbed the superhalo, has been reported at energies above $\sim1\,\text{keV}$ \citep{Lin1995}. 
The core and halo populations can be adequately described by bi-Maxwellian distributions, to account for temperature anisotropies, as in \cite{Pilipp1987} and later work. However, more recent studies have chosen to model the halo as a bi-kappa function \citep[e.g.,][]{Maksimovic2005,Stverak2009}.  
The strahl population is more difficult to characterise. Some authors \citep{Maksimovic2005,Tao2016} have calculated numerical moments directly from isolated strahl populations. However, the methods for isolating strahl velocity distribution functions (VDFs) can be limited as they are derived from subtracting distributions from different pitch angle bins, over a limited energy range, leading to potentially large uncertainty in the moment. These strahl moments are also subject to assumptions about the extent of the strahl in pitch angle, which can only be estimated to within the angular width of a given measurement pitch angle bin. Alternatively, fitting a model function to isolate strahl components \citep[as in][where a truncated kappa function was fitted]{Stverak2009} may circumvent the issue of energy cut-offs. However, any model functions used are rather ad-hoc below the typical energy at which the core/halo populations begin to dominate the strahl.

Moreover, the origins of the distinct strahl and halo electron populations in the solar wind are not well understood. 
Evidence has been found   that a suprathermal tail can exist in the solar wind using exospheric models \citep[e.g.,][]{Lie1997} 
 which ultimately require a seed suprathermal electron population to exist in the corona. \cite{Pierrard1999} use in situ electron VDF measurements from WIND to provide boundary conditions to their model of electron VDFs which originate at 4\,R$_{\odot}$. 
 Their results suggest  that an electron VDF which includes suprathermal electrons at 1\,AU must correspond to one which also included suprathermal electrons in the corona. The relative strength of the suprathermal tail is predicted to be considerably weaker in the corona than at 1\,AU, and the effect of coulomb collisions in influencing the distribution for slow wind electrons is predicted to be more significant than for fast wind. 
 
Using a Kappa function to model the ionising electron population in the corona, \cite{Ko1996} simulated the charge state distributions of coronal ions given different core electron temperatures and  values of $\kappa$. They predict, based on solar wind oxygen and carbon ionisation measurements, a weak suprathermal tail ($\kappa \geq 5$) should exist in the corona. At such  levels it is not expected that the influence of collisions with these electrons on ionisation equilibrium would be very significant. In  a related study,
\cite{Esser2000}
offered an explanation of  the discrepancy between remote spectral estimates of coronal electron temperature and the freeze-in temperatures measured in situ by invoking additional ionisation by suprathermal electrons. They 
 claimed that  the sensitivity of the dominant charge state ratios to $T_h/T_c$ (the ratio of modelled electron halo temperature to the core temperature) varied strongly based on species, with $\text{O}^{7{+}}{\negmedspace/}\text{O}^{6{+}}$ proving most sensitive. Different halo temperatures were thus 
 thought  to be necessary to meet the observed ionisation states for different ions in situ. 

In contrast, \cite{Laming2004} 
proposed an explanation of  the above compositional-spectral temperature discrepancy via extra heating of the coronal thermal electrons  by lower hybrid waves, in place of suprathermal electrons. The author notes that remote estimates of coronal temperature using O VI diagnostics should be sensitive to suprathermal influence. However, these lines do not appear to show evidence of this in practice, casting doubt on the predictions of the existence of a significant suprathermal electron population in the corona.

\cite{Che2014} present a model for halo formation in the corona via a dual-stream instability process which is related to nanoflares. As in the work of \cite{Lin1997}, they postulate that nanoflares  accelerate electrons in the coronal base to beams with energies on the order of keV. These beamed electrons then travel upwards in the corona, where they trigger a two-stream instability with the thermal electron population. This results in a  redistribution of energy, as discussed in \cite{Che2014a},
involving a transfer of energy from the nanoflare-triggered electron beam to the core electron
population, and the ultimate formation of an isotropic electron halo population. 
Modelling both as Maxwellians, the core-halo temperature ratio then obeys the relation:
\begin{equation}
\label{eq:che}
\frac{T_h}{T_c} \approx  \frac{n_c}{n_h}\frac{1-C_T}{C_T}+4
\end{equation}
where $T_h$ and $T_c$ are the halo and core temperatures respectively, $n_h$ and $n_c$ are the halo and core densities, and $C_T$ is the fraction of kinetic energy which is transferred to the core electrons. For values of $C_T$ approaching 1, this describes a proportionality between the core and halo electron temperatures in the corona. 
The authors argue that this feature is preserved out to the solar wind as the coulomb collision rate is insufficient to scatter halo electrons to form a single thermal distribution before reaching the low-density region of the corona. It should also be noted that the predicted height of formation of the electron halo is 1--1.1\,$R_{\odot}$; below the ion freeze-in height of $\sim2$\,$R_{\odot}$.
This opens the possibility that the VDF of suprathermal electrons in the corona may have a relationship with the charge states of minor ions, due to their common dependence on the 
coronal  core electron temperature. As ion charge states are not influenced by dynamic processes in the solar wind, a relationship between these charge states and suprathermal electron VDFs persisting at 1\,AU would indicate that these electrons have propagated out to 1\,AU relatively unaltered themselves. 
 
 \cite{Maksimovic2005} and \cite{Stverak2009} 
 showed that the relative density of the halo population increases with heliocentric distance at the apparent expense of the strahl. They thus infer that the strahl is scattered into the halo continuously. \cite{Owens2008} estimated the degree of scattering necessary in such a case to counteract the effect of magnetic focusing during solar wind expansion and thus presented an explanation to the observed pitch angle widths of strahl. 
Modelling by \cite{Vocks2005} predicts that this scattering is caused by wave-particle interactions, notably with whistler waves.
\cite{Seough2015} put forward an alternative description involving asymmetric pitch-angle scattering of the halo caused by the relative drift between the core and halo. They predict that the strahl is the unscattered field-aligned portion of the halo which results from this asymmetry. 
Both of the above descriptions would mean that the halo and strahl can be considered to be largely scattered versions of the same population. 
Such scattering could potentially distort solar wind electron VDFs to the point at which an initial relationship with heavy ion charge states is no longer apparent at 1\,AU.

Results from a study by \cite{Hefti1999} using solar wind ion and electron observations suggest  that there is an influence from the coronal source evident in the in situ suprathermal electrons at 1\,AU.  
Combining heavy ion data from the SWICS instrument on the ACE spacecraft with electron data from the 3DP instrument on WIND, these authors reported a relationship to exist between properties of the electron suprathermal tail and the charge state ratio of oxygen. In particular, the energy content of the suprathermal tail was characterised by defining an effective suprathermal temperature (hereafter $T_{\text{eff}}$
) using the VDF derived from WIND electron measurements. This temperature is derived by differentiating the equation for a single Maxwellian distribution:  
\begin{equation}
\label{eq:teff}
 T_\text{eff} = \frac{{-}1}{k(d\ln f/{dE})} 
\end{equation}
where $k$ is the Boltzmann constant, $f$ is the electron VDF, and E is energy. For a pure Maxwellian distribution, $d\ln f/{dE}$ would be constant with energy. However, as observed suprathermal electrons do not follow a perfect Maxwellian, particularly at higher energies, this $T_{\text{eff}}$
\ calculated with observational data in fact varies with energy.

Applying this calculation for $T_{\text{eff}}$
\ at a number of energies (300, 500 and 800\,eV) to solar wind data at a boundary between two slow-fast wind transitions, a correlation between the variation of $T_{\text{eff}}$
\ (at a given energy) and $\text{O}^{7{+}}{\negmedspace/}\text{O}^{6{+}}$ was investigated by \cite{Hefti1999}. The authors restrict themselves to two periods in the solar wind observations where the wind speed has just increased significantly between streams, and find that $\text{O}^{7{+}}{\negmedspace/}\text{O}^{6{+}}$ varies similarly to $T_{
\text{eff}}$
. This leads them to conclude that suprathermal electrons at 1\,AU retain information about their coronal source as ionisation states do.
However, the study is limited to only two short ($\sim5$ day) intervals in the ACE and WIND datasets. As such, this relationship is yet to be more generally verified.

On the basis of the studies described above, one could expect that, directly above the region of solar wind formation,  a positive relationship between the energy content of the suprathermal electrons and heavy ion charge states might exist. The low-collisional nature of suprathermal electrons in the solar wind suggests the possibility that they may retain these coronal signatures, and therefore a relationship with the ions, out to 1\,AU. 
In particular, the field-aligned strahl electrons may be most likely to retain such information, as their far more rapid propagation through the heliosphere should subject them to less scattering  \citep[e.g.,][]{Owens2008}. 

In this paper, we attempt to re-examine the possible preservation of a coronal electron temperature signature in suprathermal electrons (both halo and strahl) at 1\,AU by evaluating their possible relationship with charge states of heavy ions sampled in the same streams of solar wind at L1. We first attempt this by addressing limitations of the \cite{Hefti1999} method by fitting the entire core and halo/strahl range of energies using a Maxwellian+kappa fit, and compare parameters drawn from these fits to the $\text{O}^{7{+}}{\negmedspace/}\text{O}^{6{+}}$ charge state ratio. Further, we isolate the strahl portion of the electron distribution, and take partial moments of these to test for any relationship of the strahl at 1\,AU with the electron temperature of its source.
We use the suprathermal electron parameters produced through these methods in a statistical analysis over a large dataset, to robustly explore the nature and repeatability of this possible relationship. We do so with the view that a positive relationship is indicative of an observational agreement with the description in the previous paragraph, while a negative relationship is indicative either that this description is not accurate, or the relationship has been heavily altered en route to 1\,AU, in either the corona or solar wind.

\section{Data}
\label{sec:data}

We use ion charge state data from  ACE-SWICS and electron flux data from WIND-3DP  to approximate simultaneous observations of solar wind heavy ions and suprathermal electrons as closely as possible. The time period considered covers 1998--2011, during which both satellites spent the majority of their time orbiting L1. 
Additional  magnetic field measurements are taken from the  WIND-MFI instrument. All data used in this study are available from NASA Space Physics Data Facility's CDAWeb service (https://cdaweb.sci.gsfc.nasa.gov).
The Solar Wind Ion Composition Spectrometer (SWICS) on the Advanced Composition Explorer (ACE)
measures the properties of solar wind ions.  
SWICS is designed to measure the mass and charge of common solar wind ions with masses ranging from H to Fe to determine their ionisation and isotopic states \citep{Gloeckler1998}. 
Ion charge state data are provided in the form of charge state ratios or mean charge, depending on the species, at 1-hour time resolution.

The 3-Dimensional Plasma Analyser (3DP) instrument on WIND measures 3-dimensional distributions of electrons and protons using 4 electrostatic analysers (2 per species, collectively covering 3\,eV--30\,keV), and 2 solid state telescopes which measure electron energies up to 400\,keV and protons up to 6\,MeV \citep{Lin1995}.  The electrostatic analysers allow the production of electron and proton velocity distributions as functions of look-direction. 
We derive electron distribution functions from differential electron flux spectra measured by the Electron Electrostatic Analysers; EESA-L  (${\sim}5\,\text{eV}{-}1$\,keV commonly at ${\sim}30\,\text{s}$ cadence) and EESA-H (${\sim}100\,\text{eV}{-}30\,\text{keV}$  commonly at ${\sim}98\,\text{s}$ cadence). 
Data from each analyser are available with look-directions re-binned into 8 electron pitch angle (PA) values (${\sim}15^{\circ}$, $35^{\circ}$,  $57^{\circ}$,  $80^{\circ}$,  $102^{\circ}$,  $123^{\circ}$,  $145^{\circ}$ and  $165^{\circ}$ relative to the magnetic field direction). 
We use the magnetic field vector, \textbf{B}, produced by the WIND Magnetic Field Investigation \citep{Lepping1995}
to convert the pitch angles such that they span from the direction of electron propagation along the field line which is anti-sunward (that is, the common strahl direction) to sunward.
These shall henceforth be referred to as PA bins 1 to 8, and the VDFs which are derived from the fluxes in these bins as $f_1$ to $f_8$. Bin 1 is the anti-sunward bin which will most commonly contain strahl, while bin 8 will contain strahl in the less common case of a sunward beam. 
To minimise computation time, these distributions are averaged to the same time resolution as the 1-hour SWICS heavy ion data to which the moments will be compared.

The WIND spacecraft is subject to positive charging on the order of $5-15\, \text{V}$. Estimates of spacecraft potential, $\Phi$, are available in the ``WI\_ELM2\_3DP'' dataset on CDAWeb. 
 A positive potential  provides a fixed additional energy to all detected solar wind electrons.
The potential also accelerates photoelectrons towards the spacecraft, which appear only at energies below that corresponding to $\Phi$ (to within the energy resolution of the detector). To remove the photoelectrons and correct the energies measured, we shift the energy bins down by a value equal to $\Phi$. Data from any bins which are thus assigned a negative energy are considered  photoelectrons and removed from the analysis. 
Note however that the suprathermal electron parameters calculated here in all cases concern electrons too high in energy to be contaminated by a photoelectron population. This energy range is also high compared to $\Phi$, which 
  means that the suprathermal electrons are not significantly altered by the acceleration due to the positive potential. We thus continue our analysis under the assumption that any possible inaccuracies in the reported value of $\Phi$ are insufficient to alter our ultimate conclusions.

During the chosen period 1998--2011, suitable data are sometimes sporadic due to the orbit of WIND taking it away from L1. In particular, gaps exist due to this in the data taken before 2005. We have chosen time periods where WIND spends several days at a time near L1 with which to carry out this study. Data were used only when WIND's orbital position data indicated that it was within 100\,$\text{R}_E$ of L1. This distance can be compared to the correlation length scale of the solar wind, which is typically 100\,$\text{R}_E$ or larger at 1\,AU \citep[see e.g.,][]{Wicks2009,Wicks2010}. To maximise the compatibility of the data from the two spacecraft, solar wind proton bulk velocity measurements taken from ACE-SWICS were compared with those from WIND-SWE. Cross-correlation was performed on the proton velocity data from both WIND and ACE to reveal what time-lag was present between the two spacecraft. The calculated time lags were always smaller than the 1-hour time resolution of the SWICS composition data available, and so no corrective time-shifting was performed on the data. We thus consider ACE and WIND to be sampling the same packets of solar wind for the majority of  periods used in this study, to within the resolution limits of the data. 

\section{Methodology} 
	\label{sec:method}
	
	\subsection{Charge state ratio}\label{sub:QSR}
	\begin{figure}[h]
	\begin{center}
	\includegraphics[width=19pc]{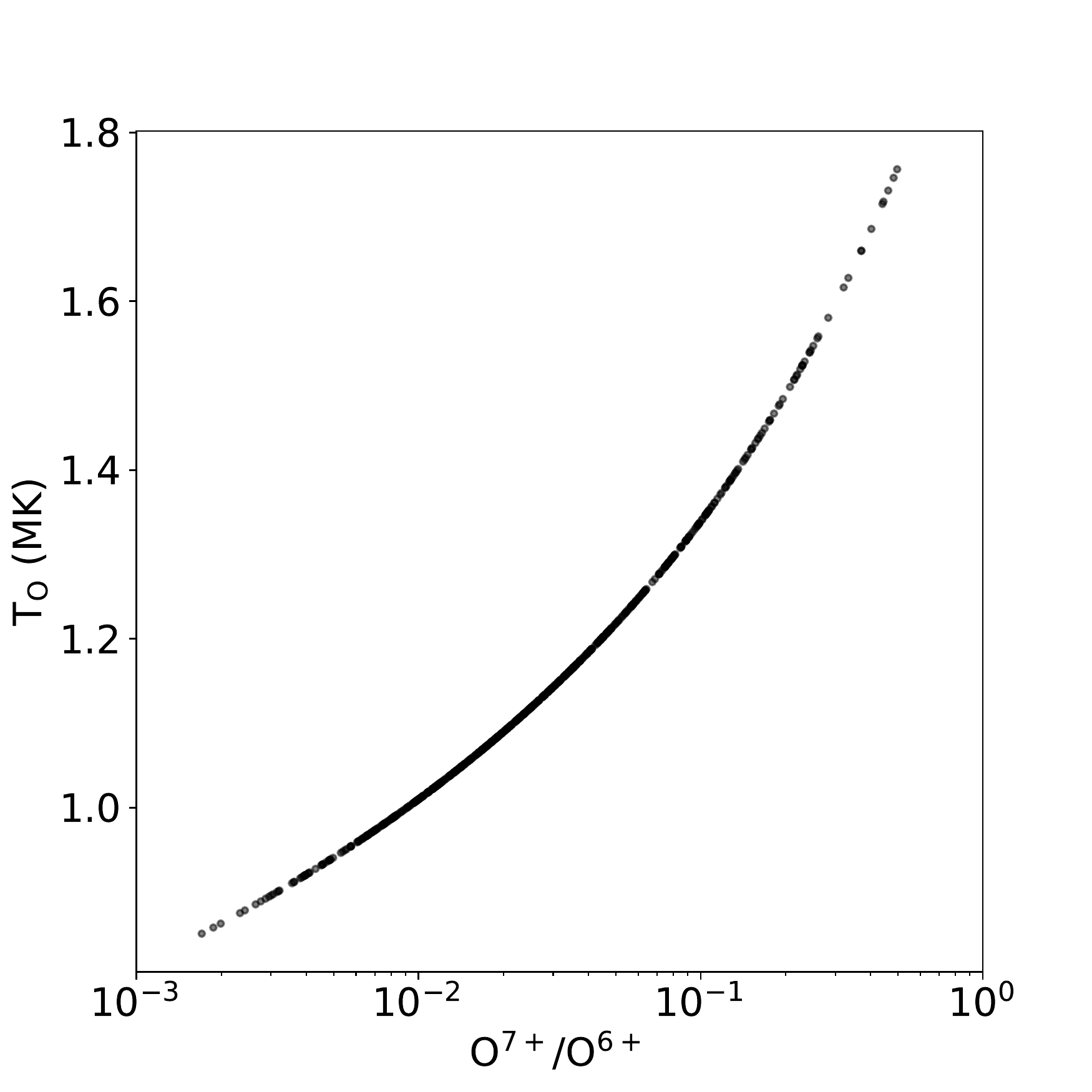}  \\
	\caption{Plot of oxygen freeze-in temperature $T_O$ against the corresponding oxygen charge state ratio $\text{O}^{7{+}}{\negmedspace/}\text{O}^{6{+}}$, from which it is calculated, taken from SWICS during the year 2007. Note that $\text{O}^{7{+}}{\negmedspace/}\text{O}^{6{+}}$ is plotted on a logarithmic scale, demonstrating that linear variations in $T_O$ correspond to order-of-magnitude variations in $\text{O}^{7{+}}{\negmedspace/}\text{O}^{6{+}}$.}\label{fig:tfreeze}
	\end{center}
	\end{figure}
	We choose the data product of oxygen charge state ratio $\text{O}^{7{+}}{\negmedspace/}\text{O}^{6{+}}$ as the primary in situ tracer of coronal temperature. 
	Figure \ref{fig:tfreeze} shows a plot of the oxygen freeze-in temperature, $T_{O}$, as calculated from SWICS measurements of $\text{O}^{7{+}}{\negmedspace/}\text{O}^{6{+}}$ collected throughout 2007, derived by solving Equation \ref{eq:ratio} for $T_e$. We do so using lookup tables of ionisation fractions 
as a function of electron temperature from the CHIANTI database \citep{Dere1997,Landi2013}, which can be rearranged to find the temperature corresponding to a given charge state ratio. 
	From the figure we see that variations in $\text{O}^{7{+}}{\negmedspace/}\text{O}^{6{+}}$ over an order of magnitude correspond to variations of $<50\%$ in the oxygen freeze-in temperature. We also note the range of $\text{O}^{7{+}}{\negmedspace/}\text{O}^{6{+}}$ observed, which approaches three orders of magnitude.
	
	We take steps to ensure that plasma associated with interplanetary coronal mass ejections (ICMEs) is excluded from our analysis. 
	To do so we follow the method of \cite{Elliott2012}; identifying as ICME times all of the intervals indicated by the Richardson and Cane list \citep{Richardson2010}, with additional time 15 hours before and 6 hours after the interval, to account for associated compressions and timing uncertainties. Any periods which fall within these criteria are not included in the analysis of subsequent sections.  
	\subsection{Core + suprathermal fits}
	\label{sub:fitmeth}
	We fit the WIND electron data to a core-halo consisting of the sum of a Maxwellian and kappa function, as was found to be suitable in \cite{Maksimovic2005} and \cite{Stverak2009}. 
	We do this in both parallel and perpendicular directions without removal of strahl electrons from the parallel VDFs. As a result, fits made parallel to the field will include both halo and strahl electrons within a single kappa function, which should ideally be used only to describe one population. The potential consequences of this for the results will be discussed in Section \ref{sec:disc}. 
	In terms of kinetic energy, $W = \frac{1}{2}m_ev^2$, the kappa function used is of the form 
	\begin{equation}\label{eq:kappa}
		f(W) = n_e\left( \frac{m_e}{2\pi\kappa W_0}\right)^{3/2}\frac{\Gamma(\kappa+1)}{\Gamma(\kappa-1/2)} \left( 1+\frac{W}{\kappa W_0}\right)^{-(\kappa+1)}
	\end{equation}
	adapted from \cite{baumjohann1997}. Here, $n_e$ is electron number density; $m_e$ is the electron mass; $W_0 = k_bT(1-3/2\kappa)$;  $\kappa$ is a dimensionless value $\geq 1.5$; and $T$ is the temperature defined by the 2nd moment of the distribution, and is independent of $\kappa$. \citep{Livadiotis2013}. This formulation can also be modified to allow for the distribution to shift up or down in energy by applying a uniform offset to $W$. 
	WIND-3DP EESA-L and EESA-H data are combined to give the full electron distribution between ${\sim}$5\,eV and 1.5\,keV; the approximate energy range spanned by the core-halo-strahl populations. Energies above this range may contain the super-halo population \citep{Lin1995}. 
	We fit the electrons to the VDFs in two pitch-angle directions separately: $f_1$, the bin closest to parallel, and $f_\perp=(f_4 + f_5)/2$, which averages the two bins either side of $90^\circ$.
	
	We attempt to fit the core and suprathermal populations independently.  For example,  we make effort to ensure that the suprathermal number density could not be decreased at the expense of an increase in the core number density during fitting. This is motivated by the premise of the study; that while the core electron population may not be expected to reach 1\,AU unaltered from its coronal state, it is more reasonable to think that the suprathermal populations might. We therefore take care not to allow influence of ``non-coronal'' distributions (the core) on our potentially ``coronal'' parameters (those which describe the suprathermals). We apply a similar method to that in \cite{Stverak2009} to achieve this. 
	We first estimate the break-point energy between the core and suprathermal populations (hereafter $E_{b}$).  This is taken to be the energy at which each population makes equal contribution to the combined VDF. For the case in which both distributions were Maxwellian, each would form a straight line in log-linear
	 space ($f$ log; $E$ linear).  The break-point would then occur when the two lines intersect.  
	 An example of where we would expect the break-point energy to lie is labelled in Fig. \ref{fig:vdfs}. We note that in the log-log space of the figure, the break can be seen as a shoulder in the distribution. 
	We find that the fitted kappa tail of the combined suprathermal distribution is very rarely smaller than $\kappa = 6$, and so at lower energies the halo is anyway closely approximated by a Maxwellian. 
	We fit the core and halo portions 
	of the VDF each to a straight line; discounting the contributions from energy bins between 40--150\,eV in order to avoid energies at which we may expect the break to lie. 
	The energy at which the fitted lines meet is then calculated, and used as the break-point energy for that VDF. The uncertainty in $E_\text{b}$, $\sigma_{\text{b}}$, is estimated from the error in the fitted parameters from each line. 
	
	Once a break-point has been found, we perform the fits for each population independently of each other, as in \cite{Maksimovic2005}. To ensure that there is no contribution of one population to the fitting of the other, the core is fitted to a Maxwellian between the limits $0 < E < (E_\text{b} - \sigma_{\text{b}})$ and the suprathermals to a kappa function within the limits $ (E_\text{b} + \sigma_{\text{b}}) < E < 1.5\,\text{keV}$. 
	This method results in a possible overestimate of the core density of approximately 2--5\% due to the halo contribution in that energy range \citep{Maksimovic2005}. 
	
	For a given pitch angle bin, the core is fitted with two parameters for the Maxwellian density ($n_\text{c}$) and temperature ($T_\text{c}$). The suprathermals are fitted with three parameters which in a typical kappa distribution represent density ($n_\text{h-s}$), temperature ($T_\text{h-s}$) and kappa ($\kappa$). 
	We have denoted these parameters as \text{h-s}, as they describe a combined suprathermal population of both halo and strahl.  
	These parameters are calculated for each bin as though they contain independent distributions. Usually we would employ bi-Maxwellian and bi-kappa distributions to produce parallel and perpendicular temperatures and one common density. However, any strahl electrons complicate this method, as they exist predominantly in the anti-sunward direction but not the sunward. The fits to the VDF for each pitch angle bin are thus assigned separate temperature and density parameters which are not constrained to be identical for all pitch angles. This means that a value of $n_\text{h-s}$ for a given direction represents the number density of the distribution were it integrated across all pitch angles as though it were isotropic. 
Attempting to derive the strahl number density with the calculation $n_s=n_{\text{h-s1}}-n_{\text{h-s}\perp}$ would then overestimate $n_s$, as the strahl is narrow in pitch angle. It is not strictly accurate, then, to describe these parameters as true temperature or density measurements of the suprathermal populations. Instead we refer to these as ``proxy'' suprathermal temperature and density (or proxy temperature and proxy density) through the remainder of this work.

	Figure \ref{fig:vdfs} (a) gives an example of the fitting method for the anti-sunward distribution, $f_1$, and the perpendicular distribution, $f_{\perp}$. The fits rapidly diverge from the data above 1\,keV as these energies are not included in the fitting process to exclude the superhalo population. For this reason these example plots are cut off at 1\,keV. The increase in $f_1$ over $f_{\perp}$ in the strahl energy range ($\sim E>100$\,eV) is clear, and we find that the fitting algorithm primarily accounts for this with an increase in $n_\text{{h-s}1}$.
	
	If the strahl is present in the solar wind at a given time, then it should be primarily contained in the kappa fit to $f_1$. The halo is thus best described using the fits to $f_8$ and $f_{4+5}$, as the fits for these are not expected to encompass strahl electrons and instead will only describe the assumed near-isotropic halo. The parameters arising from the fit to $f_1$ are a result of a combined strahl and halo population, and so do not necessarily describe either to a satisfactory degree. To test for coronal signatures carried by the strahl electrons alone, the strahl must be isolated from the halo.

	\subsection{Strahl Characterisation}\label{sub:strahlmeth}
	In the studies by \cite{Maksimovic2005}, \cite{Stverak2009} and \cite{Tao2016}, the authors all subtract some approximation of the core and halo contribution from the anti-sunward pitch angle data to isolate the strahl contribution to the VDF. We follow most closely the method of \cite{Tao2016} as their study concerns the same WIND-3DP dataset.
	
\begin{figure}[h]
	\begin{center}
	\begin{tabular}{c}
	 \includegraphics[width=20pc]{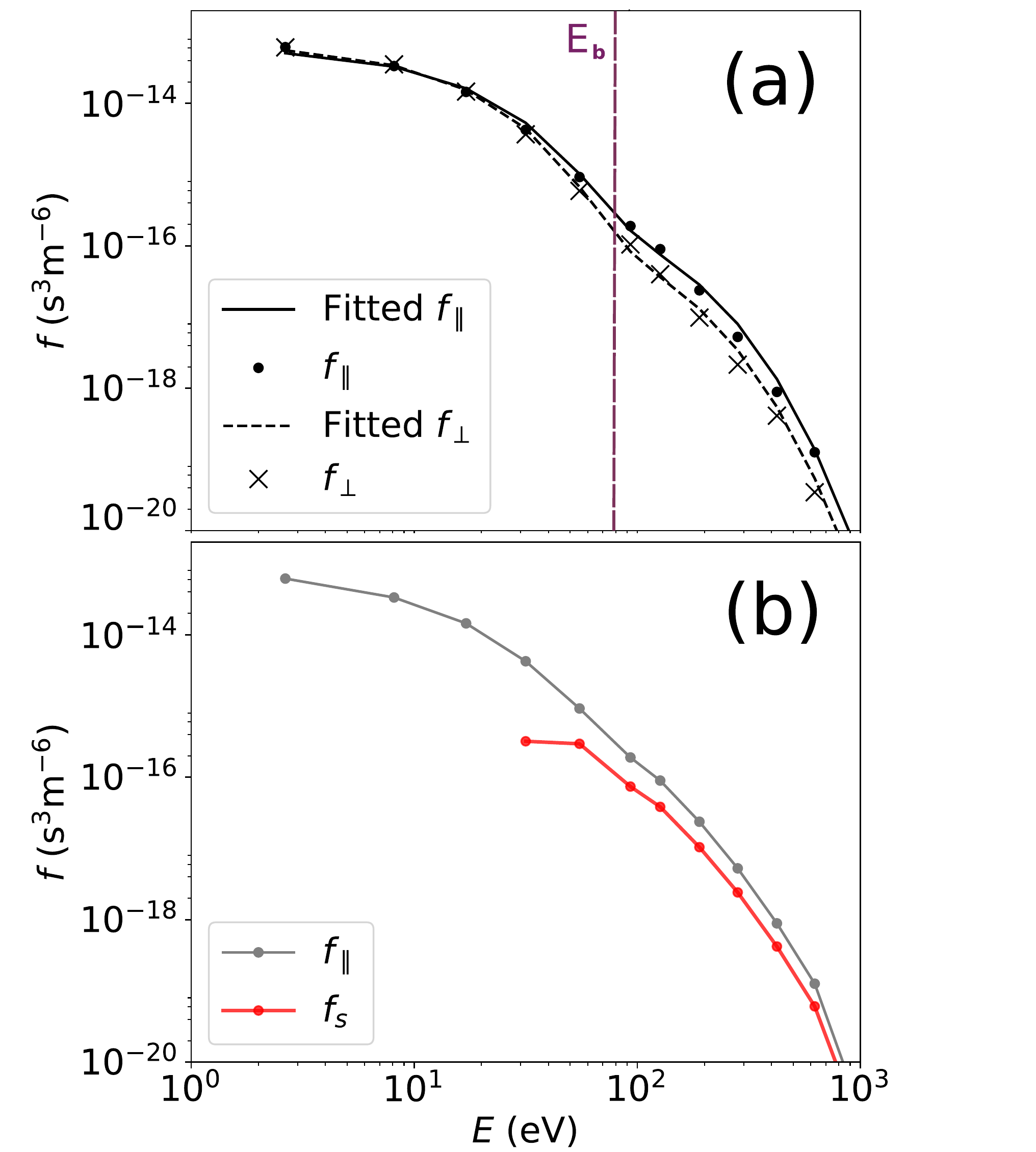} 
	\end{tabular}
	\caption{Example electron distribution functions calculated from time-averaged WIND-3DP flux data. Both (a) and (b) plot $f(E)$ on logarithmic axes. Spacecraft potential corrections have been applied to each. (a) shows data overlaid with fitted curves as described in the text. $f_1$ is enhanced over $f_{\perp}$ due to the presence of the strahl population at energies $>\sim100$\,eV. 
	An estimate of the break-point energy $E_b$ is shown in purple.  (b) shows an example $f_s$ distribution calculated as described in the text. Also shown is the corresponding $f_1$ distribution. $f_s$ makes up a significant portion of $f_1$ at energies $>100$\,eV. $f_s$ drops off rapidly below this energy, and is non-physical below 30\,eV as it is numerically negative.}\label{fig:vdfs}
	\end{center}
	\end{figure}

	The strahl angular width is assumed to be less than 45$^\circ$, and so contained entirely within PA bins 1 and 2. While \cite{Tao2016} subtract from this the mean distribution taken from $f_3$--$f_8$, taking this average does not address that these bins can not be expected to contain identical VDFs due to halo anisotropy. 
	As bins 1 and 2 are close to the parallel direction, we subtract from these the corresponding data in bins 7 and 8 
	which, in the case that a bi-kappa function models the halo accurately, should best remove the halo contribution to the near-anti-sunward VDF. The resulting VDF, 
	which we label  $f_s$,  describes the excess electrons in the anti-sunward direction which represent the strahl. 
	This method assumes that the effect of any anisotropy in the halo on $f_s$ is negligible.
	Figure \ref{fig:vdfs} (b) shows an example of $f_s$ plotted with the anti-sunward distribution $f_1$ from which it is calculated. We see that a significant portion of $f_1$ is made up of the strahl electrons in $f_s$ in the energy range $0.1\,\text{keV} \le E \le 1\,\text{keV}$. 
	Numerically, $f_s$ frequently becomes negative, and thus unphysical, at variable energies below $100\,\text{eV}$, where it is obscured by the core/halo.

	The strahl can be characterised by taking proxy-moments of $f_s$. Again following \cite{Tao2016}, we may do so by numerically integrating  
	$f_s$ within the energy range $0.1{-}1.5\,\text{keV}$. The reason for doing this is to exclude core, halo and super-halo electrons from the moment calculation. 
	Due to the hard boundary on the numerical integration, these values shall hence be refered to as ``partial moments''. This is appropriate as they do not account for all of the electrons represented by $f_s$. We distinguish these from the above proxy temperatures and  densities as they are calculated over a fixed energy range. 
	
	For the purpose of this study, we calculate only the mean energy of strahl electrons, $E_s$, through the second partial moment of the distribution: 
	\begin{equation}
	\label{eq:E_moment}
	E_s = \frac{1}{n_s} \sum_{v(100\,\text{eV})}^{v(1.5\,\text{keV})}\negmedspace\negmedspace2\pi(1-\cos{45^\circ}) f_s(v)\,\frac{1}{2}mv^4\,\Delta v
	\end{equation}
	where $2\pi(1-\cos{45^\circ})$ and $v^2$ are included as a result of the integration in spherical coordinates.
Calculating this over a constant and finite energy range should be treated with some caution, as fluctuations in the electron populations could cause the extent of the strahl to vary about these limits. 
	
	When comparing $E_s$ to $\text{O}^{7{+}}{\negmedspace/}\text{O}^{6{+}}$ we do not assign a time lag despite the strahl's more rapid propagation to 1\,AU down the magnetic field line. If the strahl is frozen into the heliospheric magnetic field, then the strahl observed at 1\,AU simultaneously to the bulk solar wind must be from the same source region in the corona. Applying a time lag would thus instead lead to comparing strahl to ion data from different source regions. We note that there is an implicit assumption that the freezing-in temperature of oxygen at the solar wind source has not changed significantly over the travel time of the oxygen ions themselves. This may be more likely to hold true for coronal hole sources than 
	it would for the slow solar wind source regions, which tend to be more chaotic and variable than the fast wind. 

	\begin{figure*}[t]
	\begin{center}
	 \includegraphics[width=42pc]{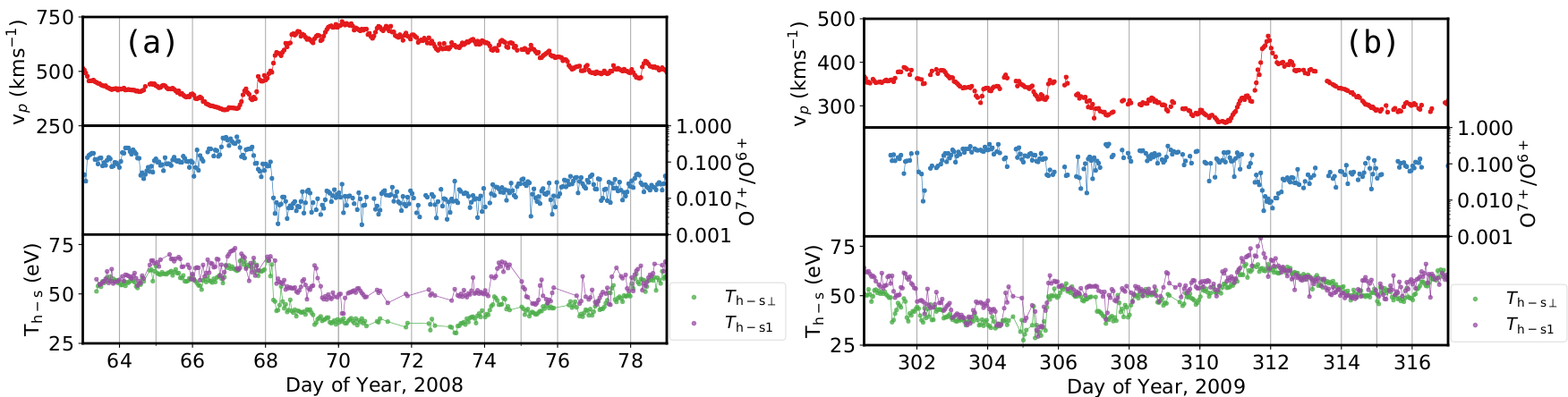}
	\caption{Time series data taken from portions of 2008 (a) and 2009 (b). Panel 1 shows solar wind proton speed as measured by the ACE spacecraft. Panel 2 plots the oxygen charge state ratio $\text{O}^{7{+}}{\negmedspace/}\text{O}^{6{+}}$ on a logarithmic scale. $\text{O}^{7{+}}{\negmedspace/}\text{O}^{6{+}}$ varies counter to $v_p$ as is expected. Panel 3 plots the pair of proxy suprathermal temperatures $T_{\text{h-s}\perp}$ and $T_{\text{h-s}1}$. These vary synchronously, with $T_{\text{h-s}1}$ tending to moderately higher values. We note that both temperatures appear to track well with $\text{O}^{7{+}}{\negmedspace/}\text{O}^{6{+}}$ in case (a), but in case (b) appear to vary oppositely with it}\label{fig:tseries}
	\end{center}
	\end{figure*}
	
\section{Results}\label{sec:res}
	We first compare time series of $\text{O}^{7{+}}{\negmedspace/}\text{O}^{6{+}}$ with proxy suprathermal temperatures $T_{\text{h-s}1}$ and $T_{\text{h-s}\perp}$. Figure \ref{fig:tseries} shows time series of solar wind bulk proton speed $v_p$ (from SWICS), oxygen charge state $\text{O}^{7{+}}{\negmedspace/}\text{O}^{6{+}}$, and temperatures $T_{\text{h-s}1}$ and $T_{\text{h-s}\perp}$ taken during 2008 (a) and 2009 (b) over $\sim$16 days. %\color{blue} 
	These time periods have been chosen to best contrast the possible relationships between these parameters, which depend on the heliospheric conditions at the time.  We observe that both $T_{\text{h-s}1}$ and $T_{\text{h-s}\perp}$ appear to vary in agreement with $\text{O}^{7{+}}{\negmedspace/}\text{O}^{6{+}}$ in (a) but in (b) vary oppositely. Viewing the data this way, it is immediately apparent that there can be no consistent tendency for our proxy suprathermal temperature to either correlate or anti-correlate with $\text{O}^{7{+}}{\negmedspace/}\text{O}^{6{+}}$. Other time periods can also be found where no apparent positive or negative relationship is clear.

\begin{figure}[h]
	\begin{center}
	 \includegraphics[width=14pc]{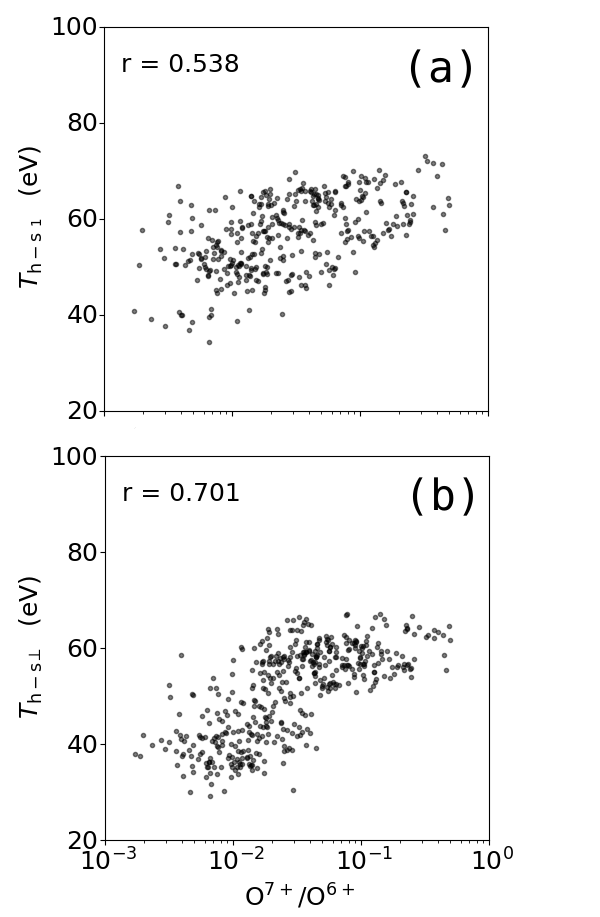} 
	\caption{Scatter plots of (a) $T_{\text{h-s}1}$ and (b) $T_{\text{h-s}\perp}$ against $\text{O}^{7{+}}{\negmedspace/}\text{O}^{6{+}}$ for Carrington rotation 2067. Pearson linear correlation coefficients are printed on the plots. Both have moderate and positive values of $r$, with $T_{\text{h-s}\perp}$ having a slightly stronger correlation. $T_{\text{h-s}1}$ is also systematically higher than $T_{\text{h-s}\perp}$ by around 5--10\,eV. }\label{fig:2067}
	\end{center}
	\end{figure}

\begin{figure}[h]
	\begin{center}
	 \includegraphics[width=14pc]{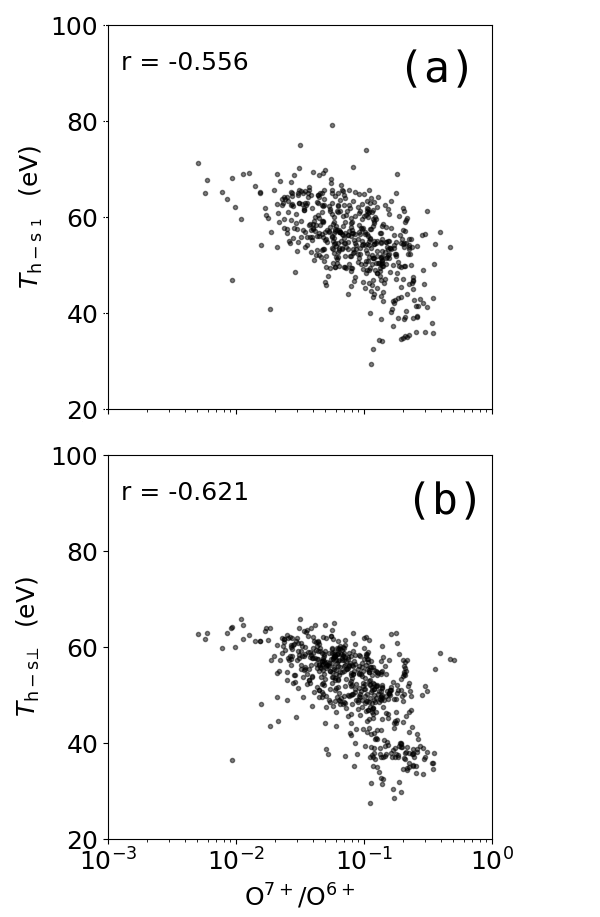} 
	\caption{Scatter plots of (a) $T_{\text{h-s}1}$ and (b) $T_{\text{h-s}\perp}$ against $\text{O}^{7{+}}{\negmedspace/}\text{O}^{6{+}}$ for Carrington rotation 2089. Pearson linear correlation coefficients are printed on the plots. Both have moderate and negative values of $r$, with $T_{\text{h-s}\perp}$ having a slightly stronger correlation. $T_{\text{h-s}1}$ is also systematically higher than $T_{\text{h-s}\perp}$ by around 5--10\,eV.}\label{fig:2089}
	\end{center}
	\end{figure}

\begin{figure}[h]
	\begin{center}
	 \includegraphics[width=14pc]{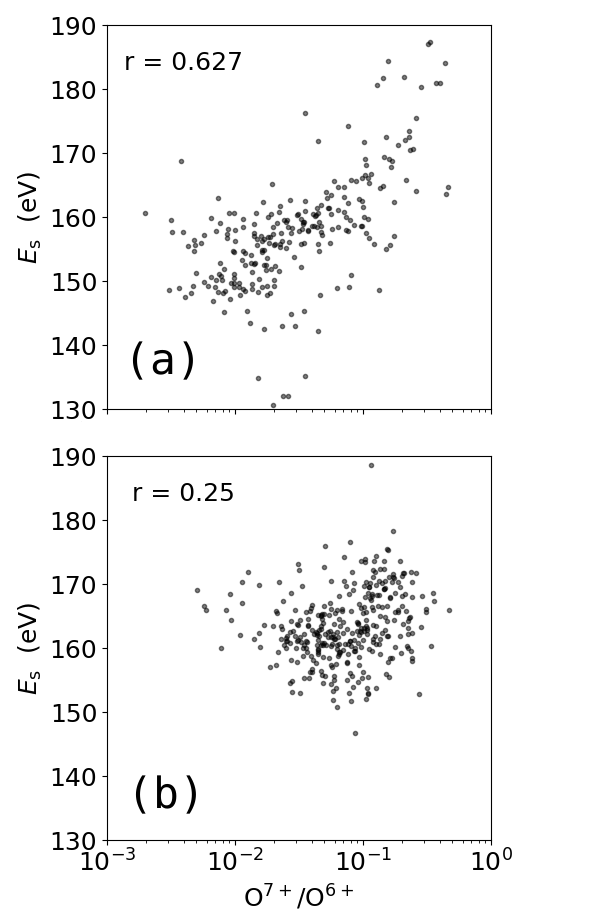} 
	\caption{Scatter plots of $E_s$ against $\text{O}^{7{+}}{\negmedspace/}\text{O}^{6{+}}$ for Carrington rotations (a) 2067 and (b) 2089.  Pearson linear correlation coefficients are printed on the plots. A weak positive relationship appears in (a) ($r=0.627$)  which is absent in (b).}\label{fig:estr}
	\end{center}
	\end{figure}

	Next, we plot proxy suprathermal temperatures against $\text{O}^{7{+}}{\negmedspace/}\text{O}^{6{+}}$ directly, to bring to light which, if any, relationship it has  with $\text{O}^{7{+}}{\negmedspace/}\text{O}^{6{+}}$. Analysing scatter plots of $T_{h-s}$ against $\text{O}^{7{+}}{\negmedspace/}\text{O}^{6{+}}$, and producing associated correlation coefficients allows for a more robust analysis of the nature of any possible relationship between the two than is possible with time series data alone. We compare data over the timescales of Carrington rotations as this allows as close to a full, instantaneous, sample of all of the solar wind in the ecliptic at 1\,AU as possible. This minimises any effects from drifting $\text{O}^{7{+}}{\negmedspace/}\text{O}^{6{+}}$ relative to $E_s$ or $T_{\text{h-s}}$ from any temporal factors on the Sun, in our correlation calculations.   
	 Figures \ref{fig:2067} and \ref{fig:2089} show the result of plotting $T_{\text{h-s}1}$ and $T_{\text{h-s}\perp}$ against $\text{O}^{7{+}}{\negmedspace/}\text{O}^{6{+}}$ for Carrington rotations 2067  (day 52{--}80, 2008; end of declining phase of the solar cycle)  and 2089 
	 (day 286{--}315, 2009; beginning of rising phase of the solar cycle),  respectively. 
	 Note that while the times in Fig. \ref{fig:tseries} (b) overlap with CR-2089, (a) does not overlap with CR-2067.  Pearson linear correlation coefficients are calculated between each temperature and $\log{_{10}(\text{O}^{7{+}}{\negmedspace/}\text{O}^{6{+}})}$. 
	 Corresponding p-values for these correlations, and those shown in Figs. \ref{fig:2089} and \ref{fig:estr}, have all been found to tend to zero, and so are not displayed on the plots themselves.  We use the logarithm of $\text{O}^{7{+}}{\negmedspace/}\text{O}^{6{+}}$ as it varies over orders of magnitude for linear changes in freeze-in temperature, as shown in the previous section. $T_{\text{h-s}\perp}$ exhibits a positive relationship 
	 ($r=0.701$)  with $\text{O}^{7{+}}{\negmedspace/}\text{O}^{6{+}}$ during CR-2067, and a negative one 
	 ($r=-0.621$)  during CR-2089. $T_{\text{h-s}1}$ varies similarly, although it has smaller magnitude in $r$ for both Carrington rotations shown. Thus for isolated time periods, and indeed whole Carrington rotations, it is possible to find both somewhat convincing positive and negative relationships. 
	This agrees with the relationships inferred from the time series data in Fig.\,\ref{fig:tseries}. We note that the $T_{\text{h-s}\perp}$-$\text{O}^{7{+}}{\negmedspace/}\text{O}^{6{+}}$ relationship shows signs of being split into a pair of populations during CR-2067; for high and low $T_{\text{h-s}\perp}$. This does not appear to be the case for $T_{\text{h-s}1}$, or for either of the relationships during CR-2089. 
	
	Turning to the electrons identified as forming the strahl populations, we now plot $E_s$ against $\text{O}^{7{+}}{\negmedspace/}\text{O}^{6{+}}$ in the same format in Figs.\,\ref{fig:estr} (a) and (b); for Carrington rotations 2067 and 2089. The former period exhibits a mild positive correlation, with $E_s$ varying over a range of approximately 20\,eV, similarly to $T_{\text{h-s}\perp}$ in the same time period. During the latter period, there is no strong positive or negative correlation, although the $E_s$ values show a similar range of variation as those in Carrington rotation 2067. We also do not observe any apparent grouping of points during CR-2067 as we did for $T_{\text{h-s}\perp}$. 
	
	\begin{figure*}[t]
	\begin{center}
\includegraphics[width=27pc]{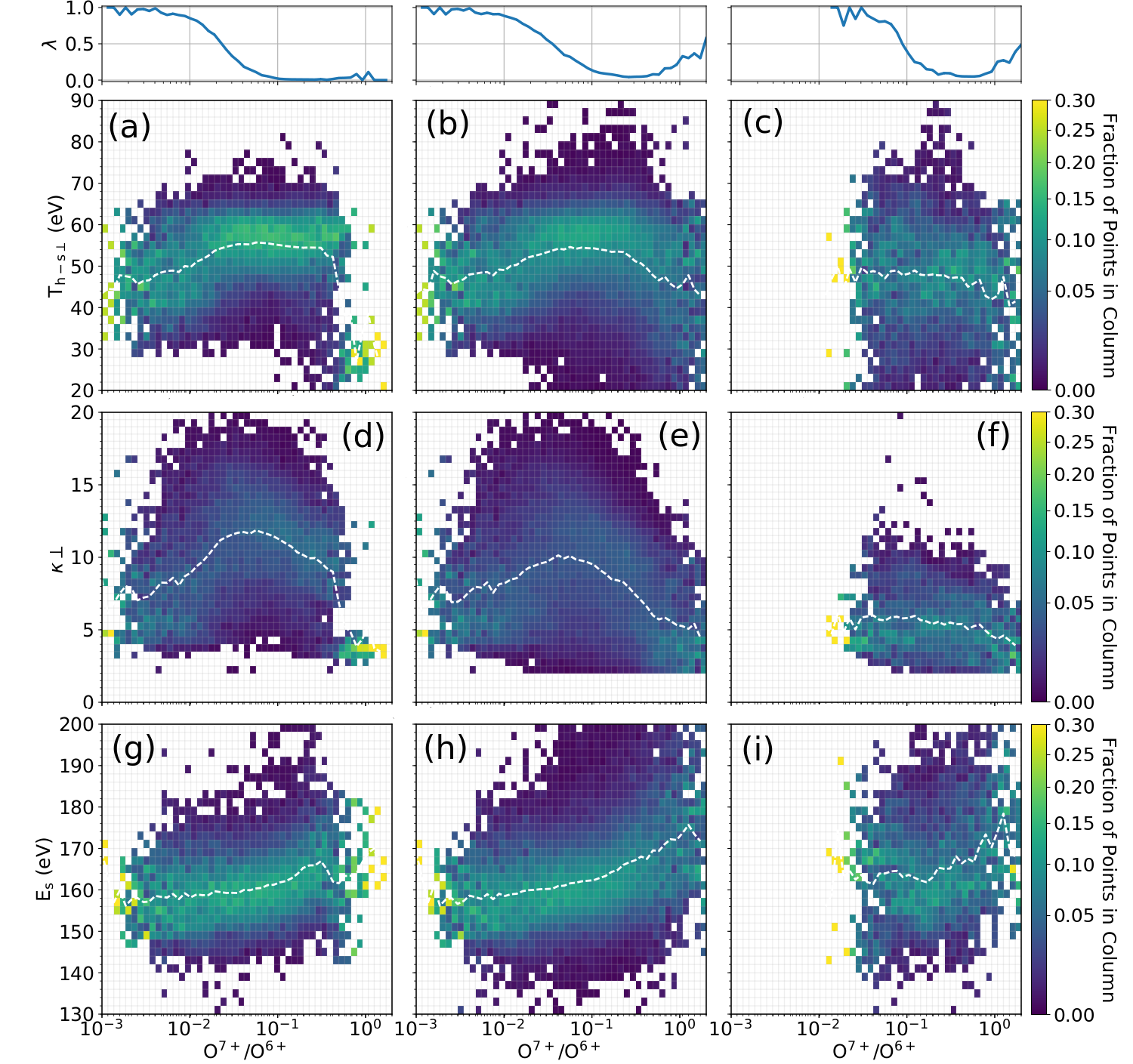}
	\caption{ Histogram plots for $T_{\text{h-s}\perp}$ (a--c), $\kappa_{\perp}$ (c--f) and $E_s$ (g--i) against $\text{O}^{7{+}}{\negmedspace/}\text{O}^{6{+}}$. (a,d,g) are composed of data taken during the lower quartile period of sunspot number, (b,e,h) the middle two, and (c,f,i) the upper. The plots are normalised for each box by the number of points in that bin of $\text{O}^{7{+}}{\negmedspace/}\text{O}^{6{+}}$. 
	For each column of plots, the small top panel above shows $\lambda$; the fraction of solar wind samples above $500 \unit{km\,s^{-1}}$ (fast solar wind) per bin of $\text{O}^{7{+}}{\negmedspace/}\text{O}^{6{+}}$.
	For $T_{\text{h-s}\perp}$, we note a weak upward trend in (a) which is not found in (c); likely as $\text{O}^{7{+}}{\negmedspace/}\text{O}^{6{+}}$ in (c) does not extend to sufficiently low values. (b) shows a similar increase in $T_{\text{h-s}\perp}$ at low $\text{O}^{7{+}}{\negmedspace/}\text{O}^{6{+}}$ values as (a) does, and also a drop-off at high $\text{O}^{7{+}}{\negmedspace/}\text{O}^{6{+}}$ values as in (c). $T_{\text{h-s}\perp}$ extends down to the pre-defined limit of halo temperature, indicating likely drop-outs of the halo at large $\text{O}^{7{+}}{\negmedspace/}\text{O}^{6{+}}$. 
	For $\kappa_{\perp}$, in both (d) and (e) we observe a broadly spread upwards trend in the low-$\text{O}^{7{+}}{\negmedspace/}\text{O}^{6{+}}$ regions ($<0.02$), which then inverts to become a downwards trend at higher values. In (f), we see only high-$\text{O}^{7{+}}{\negmedspace/}\text{O}^{6{+}}$ values and see that they decrease gradually. There is a degree of agreement between these trends and those of $T_{\text{h-s}\perp}$ above. 
	For $E_s$, a weak upward trend is found in (g), which appears to sharply fall off at higher $\text{O}^{7{+}}{\negmedspace/}\text{O}^{6{+}}$ ($>0.2$). This may be due to a lack of samples at high $\text{O}^{7{+}}{\negmedspace/}\text{O}^{6{+}}$, however. (i) also shows a weak positive trend, but this is far less smooth as (i) is made up of fewer samples than (g) or (h). (h) shows a continuous positive trend through all available $\text{O}^{7{+}}{\negmedspace/}\text{O}^{6{+}}$ samples.  The trend increases more sharply as $\text{O}^{7{+}}{\negmedspace/}\text{O}^{6{+}}$ increases.}\label{fig:histo}
	\end{center}
	\end{figure*}

	To explain the observation of both periods of positive and negative correlations between $\text{O}^{7{+}}{\negmedspace/}\text{O}^{6{+}}$ and proxies for the suprathermal temperature, we now consider the data over multiple Carrington rotations. We group the available data in time, based on the phase of the solar cycle, which we define simply by using quartiles of the monthly sunspot number, acquired from the SILSO World Data Center. 
	Our $T_{\text{h-s}\perp}$, $\kappa_{\perp}$, $E_s$ and $\text{O}^{7{+}}{\negmedspace/}\text{O}^{6{+}}$  data from the lower and upper quartile time periods (2006--2010 and 1999-2003, respectively) are shown in the left and right columns of plots in Fig.\,\ref{fig:histo}. 
	Data from the remaining two middle quartiles are shown combined in the central column. 	
	We note that there is a large portion of missing data in the upper quartile time period owing to the orbit of WIND. This period falls primarily over the time range for the sunspot maximum, leaving only around 1 year's worth of data available for that quartile in total.

	To contextualise the types of solar wind which are represented in these plots by solar wind speed as well as $\text{O}^{7{+}}{\negmedspace/}\text{O}^{6{+}}$, for each column we define $\lambda$ as the fraction of solar wind samples in each bin of  $\text{O}^{7{+}}{\negmedspace/}\text{O}^{6{+}}$ which can be considered ``fast'' ($v>500 \unit{km\,s^{-1}}$). We plot $\lambda$ against $\text{O}^{7{+}}{\negmedspace/}\text{O}^{6{+}}$ as the top panel of each column. This shows in each case that the low (high)-$\text{O}^{7{+}}{\negmedspace/}\text{O}^{6{+}}$ portions of each period contain $>80\%$ fast (slow) wind, with intermediate sections at mid-range values. We note, however, that with increasing solar activity (moving left to right) we see a trend for the transition from fast to slow to occur at higher $\text{O}^{7{+}}{\negmedspace/}\text{O}^{6{+}}$ values. Further, we attribute the up-turn in $\lambda$ at high-$\text{O}^{7{+}}{\negmedspace/}\text{O}^{6{+}}$ to be a result of the high-speed solar wind which is associated with compositionally hot active regions.

	Each main panel in Fig. \ref{fig:histo} 
	(labelled a to i) plots a derived electron parameter (on the $y$ axis)  against $\text{O}^{7{+}}{\negmedspace/}\text{O}^{6{+}}$ as a 2D histogram which has been normalised  by the number of data points in each column of $\text{O}^{7{+}}{\negmedspace/}\text{O}^{6{+}}$. In this way, the colour of each box describes the probability of measuring that value of $y$ given the corresponding $\text{O}^{7{+}}{\negmedspace/}\text{O}^{6{+}}$ value. This normalisation is applied to account for discrepancies in the number of samples at the extremes of $\text{O}^{7{+}}{\negmedspace/}\text{O}^{6{+}}$, which tends to be skewed strongly towards higher values. A dashed white line in each plot traces the weighted mean $y$ for each bin of $\text{O}^{7{+}}{\negmedspace/}\text{O}^{6{+}}$. 
	
	The data acquired during periods of quiet Sun in Fig.\,\ref{fig:histo} (a) displays a weak upwards trend for $T_{\text{h-s}\perp}$ with $\text{O}^{7{+}}{\negmedspace/}\text{O}^{6{+}}$, which climbs primarily between $\text{O}^{7{+}}{\negmedspace/}\text{O}^{6{+}}=0.002\text{--}0.02$; levelling out and falling off at $\text{O}^{7{+}}{\negmedspace/}\text{O}^{6{+}}>0.1$. This is most clearly visible in the mean line, as the spread of the data in $T_{\text{h-s}\perp}$ is very broad. The large spread in the data means that any correlation coefficient calculated from it would be very small. The histogram appears to be split into two clusters, in the bottom left and bottom right of the plot, at around $\text{O}^{7{+}}{\negmedspace/}\text{O}^{6{+}} = 0.02$. This corresponds to around $70\%$ fast solar wind.  
	
	 The data acquired near solar maximum in Fig.\,\ref{fig:histo} (c) do not exhibit an upwards trend or clustering, in contrast to the lower quartile data. However in this case the data do not extend to below $\text{O}^{7{+}}{\negmedspace/}\text{O}^{6{+}} = 0.02$, which is near the the cut-off for the clustering and upwards gradient observed in (a). This is likely due to the properties of solar wind streams which existed at these times; possibly in combination with sampling issues brought about by WIND's orbit. Nevertheless a downwards trend around higher $\text{O}^{7{+}}{\negmedspace/}\text{O}^{6{+}}$ values still seems apparent. $T_{\text{h-s}\perp}$ overall appears to be lower on average than in (a), with a wider spread that may be due to a lack of samples taken for this period. 
	
	Figure \ref{fig:histo} (b) contains the same plot as above for the remainder of the $T_{\text{h-s}\perp}$ data, covering mid-levels of activity. The left section of the plot appears to mimic the relationship found in (a), while the right mimics that found in (c). This suggests that these relationships may be dependent mostly on the availability of high and low $\text{O}^{7{+}}{\negmedspace/}\text{O}^{6{+}}$ solar wind at low-latitudes.
	
	In Fig. \ref{fig:histo} (d--f) we plot $\kappa_{\perp}$ against $\text{O}^{7{+}}{\negmedspace/}\text{O}^{6{+}}$ for different solar cycle periods. Overall this parameter exhibits far more spread than we see in  $T_{\text{h-s}\perp}$. We find that in the low-$\text{O}^{7{+}}{\negmedspace/}\text{O}^{6{+}}$ section, $\kappa_{\perp}$ increases with increasing charge state, whereas it falls with increasing charge state in the high-$\text{O}^{7{+}}{\negmedspace/}\text{O}^{6{+}}$ section. This is similar to the change in $T_{\text{h-s}\perp}$ in \ref{fig:histo} (a) and particularly (b). Further, we see $\kappa_{\perp}$ decline with $\text{O}^{7{+}}{\negmedspace/}\text{O}^{6{+}}$ in (f), as $T_{\text{h-s}\perp}$  does in (c). For all periods of the solar cycle there is a degree of agreement between these two parameters which both primarily describe the halo population. We note that this result appears to agree with \cite{Tao2016}, who reported correlation during fits to the halo population between the temperature and $\kappa$. 

	We find similar results for $E_s$ in Fig. \ref{fig:histo} (g--i) as we do  for $T_{\text{h-s}\perp}$, with some distinctions. A positive correlation with $\text{O}^{7{+}}{\negmedspace/}\text{O}^{6{+}}$ can be seen when sufficiently low values of $\text{O}^{7{+}}{\negmedspace/}\text{O}^{6{+}}$ are sampled, as is evident from the white line which illustrates the mean in (g). However, at times when these low values are not sampled in the solar wind the upward trend appears to continue. 
	The increasing trends found here are associated with a similar wide spread in underlying values to those in $T_{\text{h-s}\perp}$. 
	We find in (h)  the strongest positive trend in $E_s$ with $\text{O}^{7{+}}{\negmedspace/}\text{O}^{6{+}}$. Particularly, in the upper range of $\text{O}^{7{+}}{\negmedspace/}\text{O}^{6{+}}$ ($>0.1$)  there is a moderate  increase in $E_s$. 
	
	\linespread{1}
	\begin{figure}
	\begin{center}
	  \includegraphics[width=20pc,trim={0 1.5cm 0 1.6cm},clip]{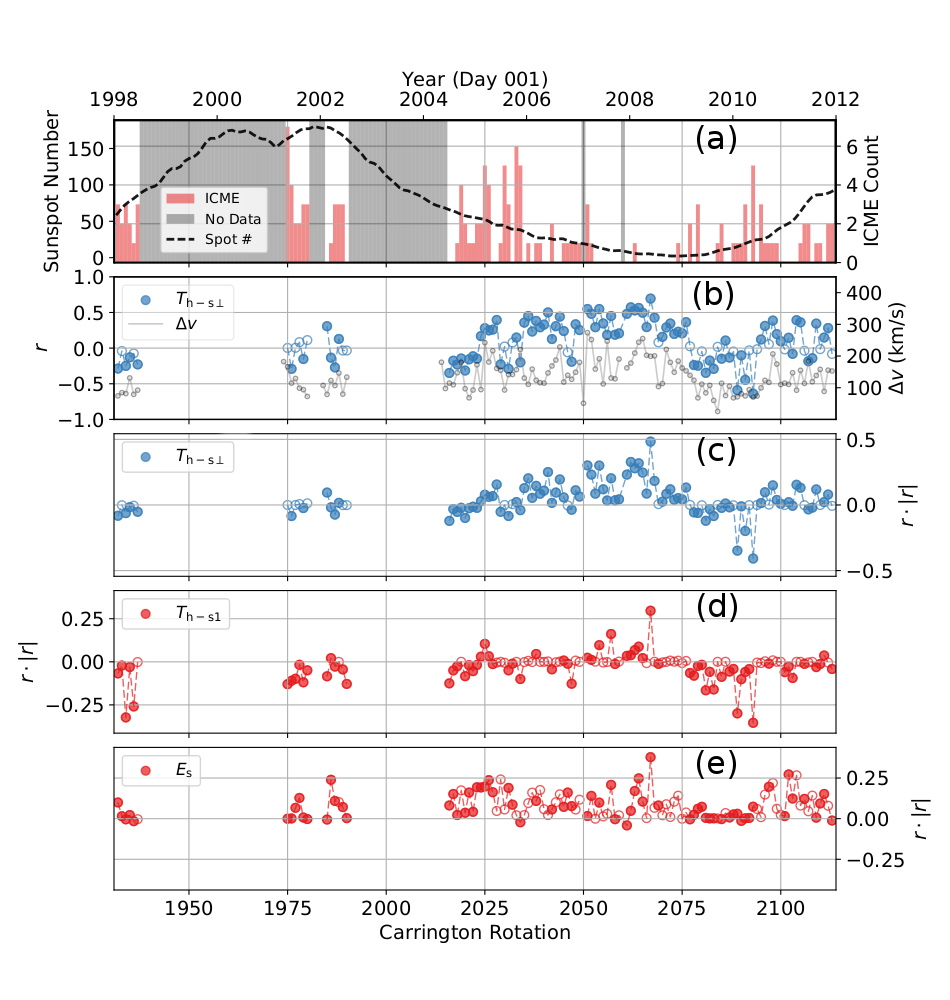}
	  \caption{Linear correlation data for suprathermal electron parameters with $\text{O}^{7{+}}{\negmedspace/}\text{O}^{6{+}}$, as calculated over single Carrington rotations against time. Also shown is supporting solar and heliospheric data. Time is shown as years on the top axis and Carrington rotation number on the bottom. All correlation data ($r$ or $r\cdot r$) are shown as a filled in point only when the corresponding p-value is $<0.05$, otherwise it is outline-only. 
	   Panel (a): Monthly sunspot number is plotted as a dashed line. A sunspot maximum followed by a minimum can be seen over the course of the observations. Also plotted is a histogram of the number of ICME detections at L1 from the Richardson and Cane list \cite{Richardson2010}. Greyed-out boxes correspond to gaps resulting from a lack of WIND electron data. The time for the minimum in ICMEs leads that of sunspot minimum by 1--2 years. This period has some agreement with that of strongest $T_{\text{h-s}\perp}$-$\text{O}^{7{+}}{\negmedspace/}\text{O}^{6{+}}$ correlation in 2007--2008. 
	  Panel (b): The left axis plots the correlation coefficient $r$ calculated for the pairing $T_{\text{h-s}\perp}$-$\text{O}^{7{+}}{\negmedspace/}\text{O}^{6{+}}$. The right axis plots the value $\Delta v$, which describes the range of the velocity data sampled as a difference of upper and lower quartiles (detailed in text). There is some apparent  tracking between these two parameters, notably in the period following 2006.
	  Panel  (c): $r\cdot r$ for the correlation coefficient between $T_{\text{h-s}\perp}$ and $\text{O}^{7{+}}{\negmedspace/}\text{O}^{6{+}}$. The fractional dependence of the two parameters on each other is clearly less that 20\% for most Carrington rotations. A notable exception to this is the period of enhanced positive correlation around the time 2007-2008. 
	  Panel (d): The same plot as Panel  (c), with $T_{\text{h-s}\perp}$ replaced by $T_{\text{h-s}1}$. The magnitude of $r\cdot r$ is almost uniformly smaller than for $T_{\text{h-s}\perp}$.
	  Panel (e): $r \cdot |r|$ for the linear correlation $r$ calculated for $E_s$ with $\text{O}^{7{+}}{\negmedspace/}\text{O}^{6{+}}$. For most Carrington rotations there is a weak, positive, relationship. There are no  notable Carrington rotations in which there is a negative relationship.}
	    \label{fig:big}
	\end{center}
	\end{figure}
	
			Returning to calculations of correlation coefficient, we repeat the calculation of $r$ for our suprathermal electron parameters against $\log{_{10}(\text{O}^{7{+}}{\negmedspace/}\text{O}^{6{+}})}$, for each Carrington rotation within the available dataset. The results of this are shown in  Fig.\,\ref{fig:big}. Panel (a) serves to contextualise the correlation data in the rest of the plot. The dashed black line shows the monthly sunspot number, showing that the full dataset spans the time of approximately one solar cycle. The first half of the data occurs around solar maximum, and the second around minimum. The histogram shows in red the occurence of interplanetary coronal mass ejections (ICMEs) detected at 1\,AU, taken from the Richardson and Cane (2010) ICME list. Greyed-out boxes show periods when absence of WIND data has prevented analysis. The period with fewest ICMEs appears to correspond to the period of maximum positive correlation for $T_{\text{h-s}\perp}$, around 2007--2008. This also coincides with the declining phase of cycle 23 indicated by the sunspot number, with a slight offset in time. Apart from this trend, there does not appear to be a direct correspondence with ICME activity and the correlation of $\text{O}^{7{+}}{\negmedspace/}\text{O}^{6{+}}$ with any of the suprathermal electron parameters on a per-Carrington rotation basis.
	
 The variation in $r$ with time is shown in  panel (b)  of Fig.\,\ref{fig:big} for $T_{\text{h-s}\perp}$ only. Filled-in points indicate correlation coefficients with a corresponding p-value of less than 0.05; a typical cut-off for signficance. 
	The square of a Pearson correlation coefficient, $r^2$, can be interpreted as the fraction of variation in the data which is described by the assumption that the two variables from which $r$ is calculated are linearly dependant. We plot $r\cdot |r|$ for the correlation of $T_{\text{h-s}\perp}$ ($T_{\text{h-s}1}$) with $\text{O}^{7{+}}{\negmedspace/}\text{O}^{6{+}}$ in panel (c) (panel (d))  of Fig.\,\ref{fig:big}. This value expresses the value of $r^2$  between the parameters, while still preserving the sign of $r$. With this parameter we easily observe that, for the majority of Carrington rotations, there is very little dependence of halo temperature on $\text{O}^{7{+}}{\negmedspace/}\text{O}^{6{+}}$, as $r^2$ rarely exceeds 25\%. 
	Panel (c) shows $r\cdot |r|$ exceeds a positive correlation with dependency of 25\%  during some Carrington rotations in 2007--2008; the period of fewest ICMEs noted above. 
	
\begin{figure}
	\begin{center}
	\includegraphics[width = 18pc]{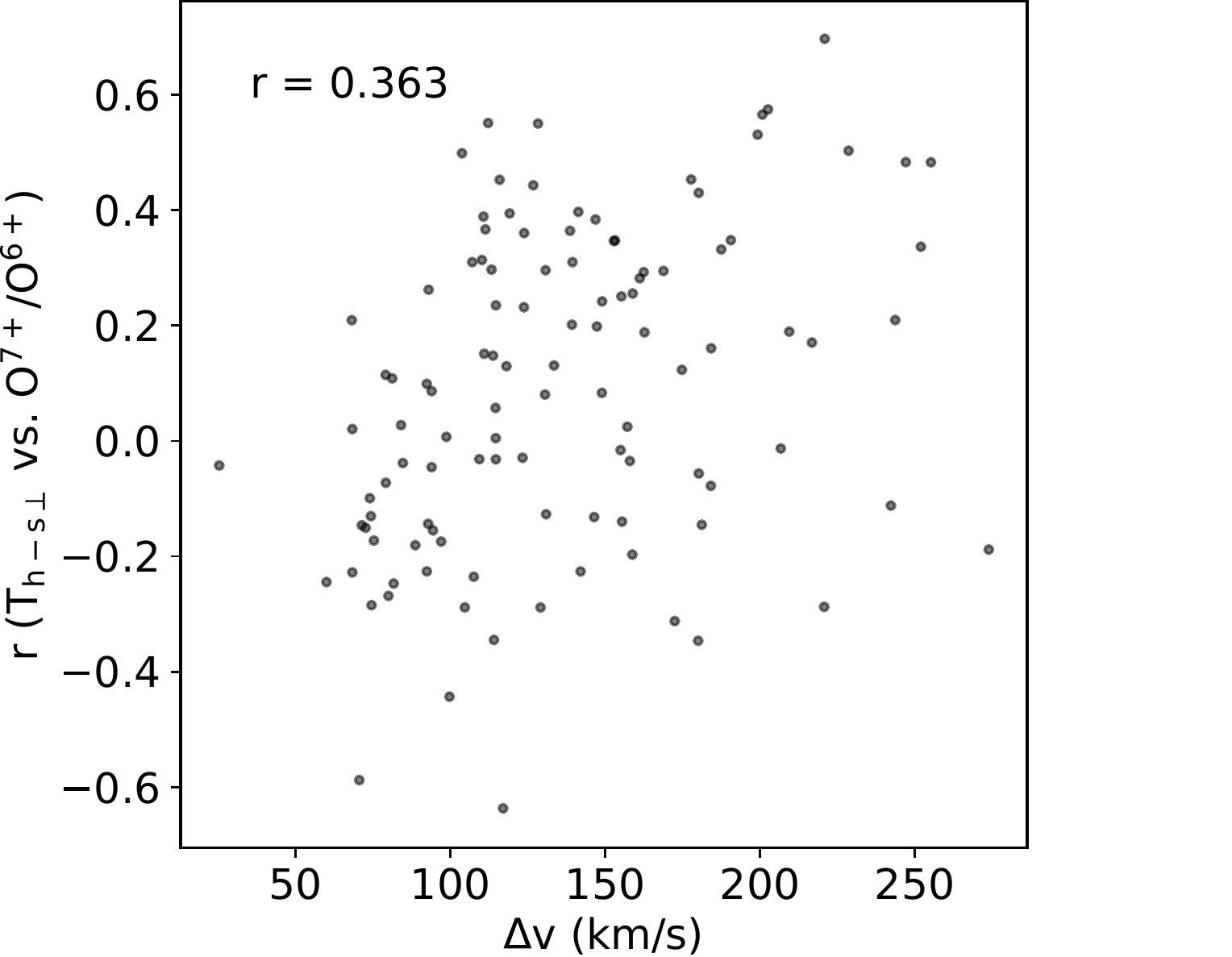}
	\caption{ Plot of the correlation coefficient $r$, calculated between $T_{\text{h-s}\perp}$ and  $\text{O}^{7{+}}{\negmedspace/}\text{O}^{6{+}}$ for each Carrington rotation, against the measure of spread in velocity $\Delta v$. There is a weak positive trend between the two.}\label{fig:vcor}
	\end{center}
	\end{figure}

	To quantify the extent to which a full sample of the available solar wind conditions have been captured for a given Carrington rotation, we define $\Delta v$ as the lower quartile value of  the solar wind speed subtracted from the upper. This provides a description of the range of velocities covered by the data, which will be smaller when the solar wind exhibits less diversity in its streams, or when a portion of the data corresponding to one velocity regime is missing. To test if there is a relationship between the degree of correlation and $\Delta v$, we plot the two directly against each other in Fig. \ref{fig:vcor}. Any apparent tracking in Panel (b) only amounts to a small correlation of 0.363.  The degree of correlation between perpendicular suprathermal temperature and $\text{O}^{7{+}}{\negmedspace/}\text{O}^{6{+}}$ is not very sensitive to the diversity of available wind speed. 
	
	Panel (e) of Fig.\,\ref{fig:big} shows $r \cdot |r|$ calculated for the correlation coefficient of $E_s$ with $\text{O}^{7{+}}{\negmedspace/}\text{O}^{6{+}}$. As with the other parameters, any relationship represented by these values of $r$ is very weak, as $r^2$ never  exceeds 0.5,  and the values are typically smaller than even those for $T_{\text{h-s}\perp}$. The most extended period where $r$ is positive appears to fall between 2004  and 2008; the declining phase of cycle 23, which is slightly longer than the extended positive period for $T_{\text{h-s}\perp}$. There is no comparable period of negative correlation, although there is a period of extended near-zero correlation which appears to correspond to the period of most negative correlation for both $T_{\text{h-s}\perp}$ and $T_{\text{h-s}\parallel}$ around the rising phase of cycle 24.

\section{Discussion}\label{sec:disc}
\subsection{Coronal temperature signatures at 1\,AU} 
	We first note that all correlation coefficients and apparent trends between $\text{O}^{7{+}}{\negmedspace/}\text{O}^{6{+}}$ and derived suprathermal electron properties  in the data which have been shown in Section \ref{sec:res} only imply, at best, weak relationships. Correlation coefficients which accompany the scatter plots in Figs.\,\ref{fig:2067}-\ref{fig:estr} and feature in the long-term analysis of Fig.\,\ref{fig:big} only correspond to values of $r^2$ which rarely exceed 10\%. As described in Section \ref{sec:res}, this value describes the fraction of variation in the data which can be explained by the two sharing a linear relationship. Likewise, while there is frequently a positive trend in the mean lines in the histograms in Fig.\,\ref{fig:histo}, the large spread in the data is indicative of the weakness of the overall increasing trend. Caution must be used when trying to explain or draw conclusions from such weak correlations, but perhaps more reasonably we can attempt to explain the weakness of the correlations themselves, and the variation therein. The weakness of the $T_{h-s}$ and $E_s$ relationships with $\text{O}^{7{+}}{\negmedspace/}\text{O}^{6{+}}$ could suggest that the suprathermal electron populations have lost almost all characteristics relating to coronal temperature signatures before they reach 1\,AU. Alternatively, this could mean that these signatures are not set in the corona in the way predicted in Section \ref{sec:intro}, i.e.,\, with more energetic suprathermal electron populations being formed in regions with higher core electron temperature, at the correct height to map to the oxygen freeze-in height. In this section we explore how the evidence may be interpreted in each case.
	 
The following discussion focuses on $T_{\text{h-s}\perp}$, which we believe is a better represention of the halo temperature than $T_{\text{h-s}1}$, while our parameter $E_s$ will be used to provide information about the isolated strahl. 
The observation in Fig.\,\ref{fig:2067}, and in later plots, that $T_{\text{h-s}1}$ tends to be greater than $T_{\text{h-s}\perp}$, is most likely due to the presence of the strahl electrons in $f_1$. 
The strahl is well-described as a beam of electrons. The effect of adding such a beam, with a relative velocity drift, to the halo is to create a new distribution which is enhanced at the energies around the beam energy.
Thus a kappa fit to this distribution returns a temperature which is enhanced over 
that for the halo alone, as long as the central strahl energy is sufficiently displaced from the central halo energy. The size of the temperature increase depends strongly on the number density of the strahl. Given these complications in interpreting $T_{\text{h-s}1}$, we do not consider it further in this section. 

\subsection{Explanations for weakness of coronal signatures}
A popular model for the formation of the core-halo-strahl feature is that the halo is formed by pitch angle scattering of strahl electrons by whistler waves,
which is balanced by magnetic focusing to maintain the field-aligned strahl \citep{Owens2008}. The scattering
and refocusing processes can occur continuously during propagation, and so electrons which arrive at 1\,AU as
either part of the strahl or halo populations could have been subject to scattering events several times during the course of
their propagation. Alternatively, \cite{Seough2015} suggest the strahl population may be
expected to have been subject to far less scattering than the halo by the time it reaches 1 AU. 
In both cases, the halo and strahl electrons are predicted to originate from the same population.
This appears to be the case within the limits of our measurements, to the extent that the two appear to both be very weakly correlated with $\text{O}^{7{+}}{\negmedspace/}\text{O}^{6{+}}$ by the time they reach 1\,AU. We explore the $T_{\text{h-s}\perp}$ and $E_s$ relationships with $\text{O}^{7{+}}{\negmedspace/}\text{O}^{6{+}}$, attempting to find evidence as to whether their state at 1\,AU is a result of an initially weak relationship, or an initially strong relationship weakened by in situ processing. 

The time period from which Fig.\,\ref{fig:histo} (a) is drawn, 2006--2010, includes the declining phase of solar cycle 23. This period has previously been found to feature highly persistent, low-latitude, coronal holes \citep{Gibson2009,Mursula2016}. 
The solar wind from such coronal holes is likely to contain the very low $\text{O}^{7{+}}{\negmedspace/}\text{O}^{6{+}}$ values which are evident in Fig.\,\ref{fig:histo} (a). 
It appears that the weak upwards trend which we observe is due to these exceptionally low $T_{\text{h-s}\perp}$ and $\text{O}^{7{+}}{\negmedspace/}\text{O}^{6{+}}$ measurements;  primarily as they form a cluster of points which contrast with the main  population at higher $\text{O}^{7{+}}{\negmedspace/}\text{O}^{6{+}}$. This is also the case for the upwards trend in (b). 
Further, this  lower population exhibits its own self-contained gradual increase in $T_{\text{h-s}\perp}$ with $\text{O}^{7{+}}{\negmedspace/}\text{O}^{6{+}}$ which is not seen for the higher values, in which there is a gradual decrease. Perhaps only wind from the coronal hole-proper, and not these transitional regions, preserves an initial coronal temperature signature in $T_{\text{h-s}\perp}$. 
This could be due to differences in the freezing-in process in these transitional regions, or due to differences in processing which occur in the solar wind as these regions develop stream interactions.

Interestingly, the cut-off between the two distinct regions in Fig.\,\ref{fig:histo} (a) seems to be at about $\text{O}^{7{+}}{\negmedspace/}\text{O}^{6{+}} \sim 0.02$; a far smaller value than those previously found to distinguish coronal hole from non-coronal hole solar wind streams \citep{Zurbuchen2002,Zhao2011}. 
By that measure, this population falls within the high extremes of coronal hole wind charge state, and so likely does not include many samples from the trailing edges of coronal hole wind streams, across which $\text{O}^{7{+}}{\negmedspace/}\text{O}^{6{+}}$ gradually increases  from typical fast to typical slow solar wind values. 

			Charge state data are available in the ACE-SWICS dataset for elements other than oxygen. These include carbon charge state ratios $\text{C}^{6{+}}{\negmedspace/}\text{C}^{5{+}}$ and $\text{C}^{6{+}}{\negmedspace/}\text{C}^{4{+}}$. Each of these can provide an estimate of coronal temperature at a different freeze-in height from oxygen. \cite{Landi2015} modeled coronal charge state evolution including ionisation both by collisional and photoionisation processes. They found that the resulting solar wind value of $\text{O}^{7{+}}{\negmedspace/}\text{O}^{6{+}}$ is likely more susceptible to photoionisation than either of the above carbon charge states.  While initial comparisons with the results which have been covered in Section \ref{sec:res} appeared very similar for  $\text{C}^{6{+}}{\negmedspace/}\text{C}^{5{+}}$ and $\text{C}^{6{+}}{\negmedspace/}\text{C}^{4{+}}$ to $\text{O}^{7{+}}{\negmedspace/}\text{O}^{6{+}}$, these have not been studied further at present. Interesting future work would compare the similarities and differences in the relationships for these ions with those discussed in this study for oxygen. 

We can compare the increase in the mean $T_{\text{h-s}\perp}$ value in the low-$\text{O}^{7{+}}{\negmedspace/}\text{O}^{6{+}}$ regions of Figs.\,\ref{fig:histo} (a) and (b) to a best-guess expected increase. Using the results of the freeze-in temperature calculations shown in Fig.\,\ref{fig:tfreeze}, given the increase in $\text{O}^{7{+}}{\negmedspace/}\text{O}^{6{+}}$ from around 0.002--0.02, we can predict an increase of around 25\% in $T_\text{O}$ in the corona. The expected core-halo relationship from \cite{Che2014} shown in Equation \ref{eq:che} then suggests an increase of 25\% should also appear in $T_{\text{h-s}\perp}$, should it be preserved out to 1\,AU. The increases in the mean $T_{\text{h-s}\perp}$ in these regions in (a) and (b) appear to be around 20\%, showing reasonable agreement with the prediction. 
This implies that there may be an underlying relationship betweeen $T_{\text{h-s}\perp}$ and $\text{O}^{7{+}}{\negmedspace/}\text{O}^{6{+}}$ which for low-$\text{O}^{7{+}}{\negmedspace/}\text{O}^{6{+}}$ wind has been smeared-out in a mostly random fashion,  either in the corona itself or, by processing in the solar wind.

In the high-$\text{O}^{7{+}}{\negmedspace/}\text{O}^{6{+}}$ regions of Figs.\,\ref{fig:histo} (a) and (b) we observe a downwards trend of $T_{\text{h-s}\perp}$ with $\text{O}^{7{+}}{\negmedspace/}\text{O}^{6{+}}$. This is counter to the expected relationship, and cannot be explained as a simple spreading out of $T_{\text{h-s}\perp}$ values. We note that high-$\text{O}^{7{+}}{\negmedspace/}\text{O}^{6{+}}$ values should generally correspond to the sources of the  slow solar wind which is typically more prone to fluctuations which can alter electron distributions. The lowering in $T_{\text{h-s}\perp}$, when compared visually to corresponding $f_\perp$ electron distributions, can be understood as the halo temperature approaching the core temperature. 
The downward trend in $\text{O}^{7{+}}{\negmedspace/}\text{O}^{6{+}}$ could then show that in the most high-$\text{O}^{7{+}}{\negmedspace/}\text{O}^{6{+}}$ slow solar wind, the halo is more prone to thermalising with the core at some point between its initial formation in the corona and its propagation to 1\,AU. This then fully erases any presumed positive relationship between $T_{\text{h-s}\perp}$ and $\text{O}^{7{+}}{\negmedspace/}\text{O}^{6{+}}$ when measured in situ at 1\,AU. 

In Fig.\,\ref{fig:histo} (g--i) $E_s$ increases with respect to $\text{O}^{7{+}}{\negmedspace/}\text{O}^{6{+}}$  differently to $T_{\text{h-s}\perp}$, in that it does so continuously, while $T_{\text{h-s}\perp}$ appears to form clusters.  
 To a small extent  we see the rise in mean $E_s$ increase in rate with increasing $\text{O}^{7{+}}{\negmedspace/}\text{O}^{6{+}}$. 
Under the presumption of an initial positive relationship between suprathermal temperature and $\text{O}^{7{+}}{\negmedspace/}\text{O}^{6{+}}$ set in the corona, for all values of $\text{O}^{7{+}}{\negmedspace/}\text{O}^{6{+}}$, this can be viewed as $T_{\text{h-s}\perp}$ entirely losing this relationship in high-$\text{O}^{7{+}}{\negmedspace/}\text{O}^{6{+}}$ solar wind en route to L1, while $E_s$ preserves it. This is as the clustering for the halo would have to develop during transit of the solar wind to 1\,AU, if we assume the strahl and halo are of common origin, as described in Section \ref{sec:intro}. 
Such an occurence is possible given the strahl's potential to reach 1\,AU far more rapidly than the halo, which propagates out with the bulk solar wind.
Alternatively, the partitioning in $T_{\text{h-s}\perp}$ could be caused by a change in freeze-in height at the corona for different source regions; changing the initial relationship with suprathermal electrons and ionisation and leading to a discontinuity in the relationship between source regions. However, this interpretation does not explain the lack of  break in $E_s$, and so we favour the former.  
The fact that the halo temperature seems to best correlate in low $\text{O}^{7{+}}{\negmedspace/}\text{O}^{6{+}}$ regions, associated with the leading edge and centres of coronal hole streams, while the strahl relationship is positive in all regions, could be explained as the halo being subject to such processing outside of these relatively unperturbed regions of fast solar wind which the strahl is not. 

In Fig.\, \ref{fig:big} we examine correlation coefficients $r$ and $r\cdot|r|$ for relationships between $T_\text{h-s}$, $E_s$ and $\text{O}^{7{+}}{\negmedspace/}\text{O}^{6{+}}$ separated by Carrington rotation. 
We find that  $T_{\text{h-s}\perp}$  shows most positive correlation with $\text{O}^{7{+}}{\negmedspace/}\text{O}^{6{+}}$ during the declining phase and subsequent minimum of solar cycle 23. The declining phase of cycle 23 is notable for the presence of extended low-latitude coronal holes; the solar wind from which is compositionally cool (low $\text{O}^{7{+}}{\negmedspace/}\text{O}^{6{+}}$). This leads to a period of extended stability in the solar wind streams during this phase. As noted in Fig. \ref{fig:histo} (a) and (b), this low-$\text{O}^{7{+}}{\negmedspace/}\text{O}^{6{+}}$ wind features a positive trend with $T_{\text{h-s}\perp}$, and so these periods may produce more positive values of $r$ because they include more wind of this type. 
As shown in Fig. \ref{fig:vcor}, the strongest positive correlations do not necessarily correspond to the broadest spread in velocity. This may be because the trend appears to invert as we move from compositionally cool to hot wind, as shown in Fig. \ref{fig:histo} (a) and (b). Calculating a correlation coefficient over the entire spread of $\text{O}^{7{+}}{\negmedspace/}\text{O}^{6{+}}$ may thus result in lower correlations because of this.

As we see evident in Fig. \ref{fig:histo} a tendency for distinct trends to exist in compositionally cool (fast) and hot (slow) solar wind,  it would be of interest to measure the correlation coefficients for solar wind data collected within isolated fast or slow streams. 
 In this case we refer specifically to data from individual streams; as opposed to combining data from multiple fast or slow streams.  Doing so would help to ensure that correlations are being calculated for ions which were frozen-into their charge states at comparable heights in the corona, as they are more likely to have originated from the same region on the Sun, which would not necessarily be the case if we were to combine data from multiple streams of wind.  Based on the low and high-$\text{O}^{7{+}}{\negmedspace/}\text{O}^{6{+}}$ clusters in Figs.\, \ref{fig:histo} (a--c), we may expect that fast streams will produce a mildly positive correlation coefficient between $\text{O}^{7{+}}{\negmedspace/}\text{O}^{6{ +}}$ and $T_{\text{h-s}\perp}$, while slow streams would likely be closer to zero or negative.  
 This would be an interesting topic for future study.   

\subsection{ICME effects on suprathermal electrons}
We can also contextualise periods of positive correlation with the ICME histogram data. We note that the strongest period of $T_{\text{h-s}\perp}$ correlation occurs between 2007 and 2008, when there is a clear lack of ICMEs detected, towards the end of the declining phase of cycle 23. This complements the above point that we observe most positive correlation when we are able to sample
stable solar wind streams  which are relatively uninterrupted by transients. 

Alternatively, we can consider the possibility of ICMEs directly affecting suprathermal electron distributions upstream of the observer before reaching L1. Although we have taken steps to remove the in situ ICME data from our dataset, suprathermal electrons propagate along the magnetic field line to 1\,AU more rapidly than the bulk solar wind, or the majority of ICMEs. As such, strahl (and indeed halo, if this population results from in situ scattering of strahl) electrons which precede an ICME at 1\,AU could have been affected upstream of the observer by the ICME  through, for example, acceleration by the shock front. 
CME eruptions would also likely alter the initialisation of the relationship between ionisation states and suprathermal electrons predicted for the corona in Section \ref{sec:intro}. 
If suprathermal electrons are accelerated by ICME shocks in the corona in a similar manner to suprathermal ions \citep[e.g.,][]{Kahler2014,Ding2015}, then this would represent a severe deviation from the scenario described in Section \ref{sec:intro}. In such a case we could not expect a relationship between these electrons and ion charge state to be preserved. 
 It is thus possible that ICMEs would have an adverse effect on the probability of observing a  positive relationship at 1\,AU both through effects in the corona itself and in the solar wind. 
This is difficult to separate from the above explanation based on the spread of solar wind parameters, as there are no other large gaps in ICMEs at L1 in the time period of data included here with which we can compare. 

There is some evidence that the influence of ICMEs on suprathermal electrons is more pronounced for halo electrons than strahl.  In Fig.\,\ref{fig:big}, $E_s$ tends to have smaller $|r|$ values than $T_{\text{h-s}\perp}$, except for one period during the years  2004 and 2005. This is despite the detection of many ICMEs around this period, which we have hypothesised may be limiting the correlation levels for $T_{\text{h-s}\perp}$. 
This disagrees with the description of direct ICME influence on the suprathermal electrons, which predicts that ICMEs should have more influence over the beamed strahl electrons than the convecting halo, as the direct ICME times are removed from the convecting solar wind observations. 
Again, the disruption from standard fast and slow streams caused by ICMEs could be the cause of the difference in correlation. As we  have already noted above, a positive $T_{\text{h-s}\perp}$ relationship with $\text{O}^{7{+}}{\negmedspace/}\text{O}^{6{+}}$ relies upon samples of low-$\text{O}^{7{+}}{\negmedspace/}\text{O}^{6{+}}$ fast solar wind streams. 
To investigate and contrast ICME effects on halo compared to strahl, we intend to perform these same correlation calculations exclusively for the ICME periods which we have removed here in a future study.

\section{Conclusions}\label{sec:cons}
	We have shown that suprathermal temperature proxies, $T_{\text{h-s}\perp}$ and $T_{\text{h-s}1}$, generally exhibit only very weak correlation with $\text{O}^{7{+}}{\negmedspace/}\text{O}^{6{+}}$. 
	From our analysis in the previous section we conclude that, outside of relationships between the large-scale streams in the solar wind structure, the temperature of suprathermal electrons has very little to no residual signatures from the coronal electron temperature of its source by the time it propagates to 1 AU. 
This contrasts with the conclusions drawn by \cite{Hefti1999}, who reported that the two were related. We do not fully contradict their conclusions however, as we too find numerous subsets of data with statistically significant correlation between the suprathermal electrons and $\text{O}^{7{+}}{\negmedspace/}\text{O}^{6{+}}$. 
Likewise $E_s$, an estimate of mean strahl energy, also shows very little overall dependence on $\text{O}^{7{+}}{\negmedspace/}\text{O}^{6{+}}$. 
Both the halo, which propagates with the bulk solar wind, and the strahl, which travels rapidly down the heliospheric magnetic field, show no consistent evidence of containing a remnant signature of the electron temperature at their coronal source.  We find that in periods where there is low solar activity, fewer ICMEs and consistent fast streams, there is a greater positive correlation with $\text{O}^{7{+}}{\negmedspace/}\text{O}^{6{+}}$ for both $E_s$ and $T_{\text{h-s}\perp}$. It is likely then that in these simple configurations of the corona and solar wind, a coronal relationship is set up and partially  preserved between the suprathermal electrons and ionisation states. In the more complex states, some combination of coronal conditions (variability of freeze-in heights, ionisation processes, temporal variation of the source) and solar wind processing (increased wave activity due to CIRs, wind streams with more fluctuations, ICME influence on halo and strahl electrons) is acting to destroy this correlation before it can be observed. From this we conclude that the description in Section \ref{sec:intro} of how such a correlation between suprathermal electrons and ionisation states could come to exist is a possibility, under favourable coronal conditions. 

 We have noted many features of these relationships in Section \ref{sec:disc} while attempting to understand whether in situ processing or coronal conditions are responsible for their weakness and variability. We find that the large spread in $E_s$ and $T_{\text{h-s}\perp}$; apparent clustering into fast and slow wind; and the lack of positive correlation during periods of increased perturbation in the corona and solar wind, could each be explained by solar wind processing effects or by coronal conditions which are source-dependent. 
The one observation which appears to clearly favour the explanation of in situ processing destroying an  initially strong relationship is found when comparing the halo relationship in Fig.\,\ref{fig:histo} (b) to the strahl relationship in Fig.\,\ref{fig:histo} (e). The continued upwards trend of $E_s$ in high-$\text{O}^{7{+}}{\negmedspace/}\text{O}^{6{+}}$ solar wind which is not seen for $T_{\text{h-s}\perp}$ can be most simply explained through solar wind processing effects being more effective on the halo population than the strahl.  
We do not find any observations which exclusively favour any coronal effects as the cause for the weak correlations. However, it is important to note that this may still be the case because we have not performed analysis of any direct solar observations which would potentially reveal such effects.

		Confirming if there is indeed a coronal relationship between the halo and strahl energy content and ionisation states which is being degraded during transport to 1\,AU requires further study. One way in which this could be developed in the future would involve the upcoming ESA Solar Orbiter mission. Using composition and electron data from the spacecraft's cruise and nominal mission phases,  which will cover heliocentric distances down to below 0.3\,AU, it would be possible to test how the correlations  considered in this paper vary with distance, and  for solar wind which is still relatively pristine with respect to its coronal source region. Should we see them improve with proximity to the Sun, then this would confirm that there is an initial state created in the corona in which the energy content of suprathermal electrons is related to core electron temperature, and which is then eroded during the  transport from 0 to 1\,AU. 
\newline{}
\newline{}
The authors declare that they have no conflicts of interest.
\newline{}
\newline{}
\textbf{Acknowledgements} \newline{}
The authors thank the Wind/3DP, WIND/MFI and  ACE/SWICS teams for provision of data in this study. 
We acknowledge use of NASA/GSFC's Space Physics Data Facility's CDAWeb service to access data  (https://cdaweb.sci.gsfc.nasa.gov).
We acknowledge the use of the CHIANTI database. CHIANTI is a collaborative project involving George Mason University, the University of Michigan (USA) and the University of Cambridge (UK).
The authors are also grateful to Lynn\,B.\,Wilson\,III, Andrew Fazakerley, Deborah Baker, and Gethyn Lewis for useful discussions. 
ARM is supported by the STFC through PhD Studentship. 
CJO and RTW are supported by STFC consolidated grant to UCL/MSSL, ST/N00722/1. 
\bibliographystyle{rusnat}
 \bibliography{ARMacneil_Final}

\end{document}